\documentstyle[11pt, aasms4]{article}
\input epsf
\begin{document}
\title{Analysis of the diffuse near-IR emission from 2MASS deep integration 
data:  foregrounds vs the cosmic infrared background}

\author{S. Odenwald$^{1}$, A. Kashlinsky$^{*,2}$, J. C. Mather$^3$,
M. F. Skrutskie$^4$, R. M. Cutri$^5$\\
$^*$ To whom correspondence should be addressed\\
$^1$Raytheon  ITSS, Code 685, \\ NASA Goddard Space
Flight Center, Greenbelt, MD 20771\\
$^2$Science Systems and Applications, Inc., \\NASA Goddard Space Flight Center,
Greenbelt, MD 20771\\
$^3$Code 685, NASA Goddard Space Flight Center, Greenbelt, MD 20771\\
$^4$Dept of Astronomy, University of Virginia, Charlottesville, VA 22903\\
$^5$Infrared Processing and Analysis Center, Pasadena, California\\
}
\received{}



\def\plotone#1{\centering \leavevmode
\epsfxsize=\columnwidth \epsfbox{#1}}

\def\wisk#1{\ifmmode{#1}\else{$#1$}\fi}

\def\wm2sr {Wm$^{-2}$sr$^{-1}$ }		
\def\nw2m4sr2 {nW$^2$m$^{-4}$sr$^{-2}$\ }		
\def\nwm2sr {nWm$^{-2}$sr$^{-1}$\ }		
\def\nw2m4sr {nW$^2$m$^{-4}$sr$^{-1}$\ }
\def\Ncut {$N_{\rm cut}$\ }
\def\lt     {\wisk{<}}
\def\gt     {\wisk{>}}
\def\le     {\wisk{_<\atop^=}}
\def\ge     {\wisk{_>\atop^=}}
\def\lsim   {\wisk{_<\atop^{\sim}}}
\def\gsim   {\wisk{_>\atop^{\sim}}}
\def\kms    {\wisk{{\rm ~km~s^{-1}}}}
\def\Lsun   {\wisk{{\rm L_\odot}}}
\def\Msun   {\wisk{{\rm M_\odot}}}
\def\um     { $\mu$m\ }
\def\sig    {\wisk{\sigma}}
\def\etal   {{\sl et~al.\ }}
\def\eg	    {{\it e.g.\ }}
\def\ie     {{\it i.e.\ }}
\def\bsl    {\wisk{\backslash}}
\def\by     {\wisk{\times}}
\def\cosec {\wisk{\rm cosec}}
\def\mic {\wisk{ \mu{\rm m }}}

\def\damin   {\wisk{^\prime\ }}
\def\dasec   {\wisk{^{\prime\prime}\ }}
\def\cc      {\wisk{{\rm cm^{-3}\ }}}
\def\ddeg    {\wisk{^\circ}\ }
\def\approxeq{$\sim \over =$}
\def\abouteq{$\sim \over -$}
\def\percm{cm$^{-1}$}
\def\percmsq{cm$^{-2}$}
\def\percmcub{cm$^{-3}$}
\def\perhz{Hz$^{-1}$}
\def\perpc{$\rm pc^{-1}$}
\def\persec{s$^{-1}$}
\def\peryr{yr$^{-1}$}
\def\te{$\rm T_e$}
\def\tenup#1{10$^{#1}$}
\def\to{\wisk{\rightarrow}}
\def\thin{\thinspace}
\def\uk{$\rm \mu K$}
\def\p{\vskip 13pt}


\begin{abstract} 

This is one of two papers in which we report the detection of structure in the 
cosmic infrared background (CIB) 
between 1.25 - 2.2 \um through the use of data from the Two Micron Sky Survey 
(2MASS). This paper concentrates on data assembly, analysis and the estimate of 
the various foreground contributions; the companion paper (Kashlinsky, Odenwald, 
Mather, Skrutskie, Cutri 2002, hereafter KOMSC) presents the cosmological 
results 
for the CIB fluctuations and their implications. By using repeated observations
of a specific calibration star field, we were able to achieve integration times 
in
excess of 3900 seconds compared to the 7.8 seconds in the standard 2MASS data 
product. This yielded a point source detection limit ($3 \sigma$) of $+18.5^m$ 
in 
K$_s$ band. The resulting co-added images were processed to remove point sources 
to 
a limiting surface brightness of $+20^m$/arcsec$^2$ or 40 \nwm2sr . The 
remaining 
maps contained over 90\% of the pixels and were Fourier transformed to study the 
spatial 
structure of the diffuse background light. After removing resolved sources and 
other artifacts, we find that the power spectrum of the final images has a 
power-law distribution consistent with clustering by distant galaxies. We 
estimate 
here the contributions to this signal from Galactic foregrounds, atmospheric 
OH-glow, zodiacal light and instrument noise, all of which are small and of different slopes. 
Hence, this supports the KOMSC identification of the signal as coming from the 
CIB fluctuations produced by distant clustered galaxies. 

\end{abstract}


\section{Introduction}
The cosmic infrared background (CIB) arises 
from the accumulated emissions from galaxy populations spanning a large range 
of redshifts. The earliest epoch for the production of this background occurred 
when star formation first began, and 
contributions to the CIB have continued until the present epoch.  
Apart from the cosmic microwave background, the most common source of luminosity 
in the universe arises in galaxies whose present-day 
clustering properties  are fairly well known on scales up 
to 100$h^{-1}$Mpc. The 
CIB, being produced by clustered matter, must have fluctuations that show the 
clustered nature 
of the underlying sources of luminosity. This signature will have an angular 
correlation function (or angular power spectrum) that distinguishes it from 
other local sources of background emission such as
Zodiacal light emission, and foreground stars in the Milky Way.
 Moreover, distant contributions 
of CIB will have a different redshift, and therefore spectral color, than 
nearby 
galaxies and sources of local emission. From galaxy evolution and cluster 
evolution models  it is possible to test the predicted slopes for the 
power spectrum of the CIB against the power spectrum of the data and thereby
improve our theoretical understanding of galaxy evolution and cluster evolution
during the first few billion years of cosmic history.

The CIB is expected to be extremely faint at wavelengths shorter than
10 \um especially when seen 
against the far-brighter foregrounds contributed by local matter within the 
solar system, as well as dust and stars within the Galaxy. 
A number of investigations have 
attempted to extract the mean intensity of the CIB from ground- 
and satellite-based data (see Hauser \& Dwek 2001 for review). This has in 
nearly all instances been a complicated 
task due to a lack of detailed knowledge of the absolute brightness levels and 
their spatial variations across the sky, of the many foregrounds, 
and other non-cosmological backgrounds 
that overlay the CIB signal. 

We used 2MASS calibration data to obtain long integration times and detected the structure 
of 
the CIB in the near-IR. 
Much of our motivation for the current work stems from the recent surprising,
and mutually
consistent, discoveries of the high levels of the near-IR CIB from the
COBE/DIRBE (Dwek \& Arendt 1998, Gorjian et al. 2000, Wright and Reese 2000, Cambresy et al. 2001) and 
Japan's IRTS datasets (Matsumoto et al. 2000). Using the fluctuations method for 
the final COBE/DIRBE maps, Kashlinsky \& Odenwald (2000, hereafter KO) measured 
the 
CIB fluctuations between 1 and 5 \um\ at the
DIRBE beam scale ($\sim 0.7^\circ$) that significantly exceed predictions from
theoretical models of galaxies with only passively evolving stellar
populations.

Our presentation is divided into two parts:  this paper presents a detailed 
discussion of the data and its assembly and the various foreground contributions 
and modeling. The companion paper (Kashlinsky, Odenwald, Mather, Skrutskie \& 
Cutri 2002, hereafter KOMSC) presents the cosmological results and briefly 
touches on their implications.   

The structure of this paper is as follows:  in Section 2
we describe the 2MASS observations including their calibration, photometry
and the resulting sensitivity limits obtainable from our deep integrations. In
Section 3, we discuss the data analysis process, point source removal 
algorithms,
beam deconvolution and image de-striping. In Section 4, we evaluate the sources
of image noise, artifacts, atmospheric and instrumental effects. Section 5
is a discussion of the Zodiacal, Galactic and local universe backgrounds and
their impact upon the detected background structure in the deep-integration
study.  

\section{Observations}

\subsection{2MASS survey}

2MASS used two 1.3-m Cassegrain telescopes, one at Mt. Hopkins in the Northern 
Hemisphere, and one at CTIO in the Southern Hemisphere. Each telescope was 
equipped 
with a three-channel camera, 
capable of observing the sky simultaneously at J (1.25 \um ), 
H (1.65 \um ), and K$_s$ (2.17 \um ) at a scale of $2\dasec$ per pixel. 
As the telescopes were scanned in declination, 
individual 1.3-second sky frames were imaged on an overlapping grid by stepping 
1/6 of the array. The frames were combined by the 2MASS data pipeline software at the Infrared Processing and Analysis Center (IPAC) to form Atlas images of 
size 512 $\times$ 1024 pixels with re-sampled 1\dasec  pixels, and an effective 
integration time of 7.8 seconds per pixel. Hereafter, we will use the term 
`image' to refer to the calibrated 2MASS Atlas images which have been co-added 
to an effective integration time of 7.8 seconds. 

A limited number of standard stars had to be observed repeatedly each night, and 
for several months at a time, to establish the photometric zero-points  for the 
data. The 2MASS standard star fields were centered on stars from near infrared standard star 
catalogs (e.g. Persson et al. 1998, Casali and Hawarden 1992). 
Each calibration observation consisted of six independent scans of a calibration 
field. Each scan is a mosaic of 48 overlapping frames creating a field approximately $8.6\damin \times 1\ddeg$ with 7.8-second integration times. The scans were made in 
alternating north-south directions, each displaced 5 \dasec  in RA from the 
previous one to minimize systematic pixel effects. Repeated observations of 
calibration fields during a night at a variety of elevation angles were also 
used to develop long-term atmospheric extinction statistics.

We selected data collected at the 2MASS CTIO facility between April and August 1998 for the photometric calibration field containing the standard star P565-C. The data consisted of 2080 calibrated, 
overlapping images covering an $\sim 8.6\damin \times 1\ddeg$ swath oriented 
north-south.  The standard star region is located in the 
direction of RA(2000) = $16^h 26^m$ and Dec(2000) = $+6\ddeg  13\damin $. The 
Ecliptic and Galactic coordinates of this region are  
(243.5$^\circ$ , 27.5$^\circ$) and  (20.8$^\circ$ , 35.0$^\circ$) respectively.
This field is located in the constellation Hercules, 5 degrees north of 
the bright star $\lambda$ Ophiuchus.

\subsection{Photometry}

A detailed description of the calibration process can be found in the 2MASS 
Explanatory Supplement (Cutri et al. 2001), and we now summarize the  details 
relevant to the study of the P565-C region. The individual calibrated images were obtained  from IPAC. Each image was processed in the same way as for the 2MASS Atlas Images  using the same pipeline software and algorithms. To prepare the 2MASS images,  each 1.3-second NICMOS camera frame had been dark-subtracted, flat-fielded and  sky-illumination corrected. 
The calibration scans, consisting of 48 images obtained in a $1\ddeg$ strip in 
declination, were subjected to a rigorous quality control on a pass/fail basis. 
If a scan failed any stage of the pipeline processing, was a part of a block of 
observations that did not pass any of the photometric quality criteria, or if 
there 
was any evidence for clouds in the sky, the calibration scans were rejected. 

Prior to co-adding each 1.3-second frame to obtain the final 7.8-second
calibration image, camera pixels with poor responsivities, excessive noise, or 
affected by cosmic rays were masked off. For each "Read 1" (R1) and "Read 2-Read 1" (R2-R1) data frame in a
scan, the appropriate nightly dark frame is subtracted, and the
corresponding average "canonical" responsivity image is divided into
it. Within each scan, a series of additive sky illumination
corrections are derived by creating -trimmed averages for blocks of at
least 42 dark-subtracted, flat-fielded sky frames. The trimmed
averaging rejects any sources within the frames and yields a
measurement of residual dark-sky illumination patterns on the
detectors within each block. This so-called "sky offset" frame is then
subtracted from each input frame, resulting in a data frame ready for
source detection and combination into the final survey Atlas
Images. The background levels of the final instrumentally calibrated
frames correspond to the original sky levels. 
Before adding a frame to the output 
image, the image background was adjusted to match that of those images already 
combined into the image by removing the median of the differences at each point 
in the sky in the overlap region between the current frame and the stack of 
already summed frames. The only background compensation made during the image 
construction was to adjust the frame backgrounds by a constant to produce 
seamless co-add images. Although the NICMOS arrays have 2\dasec  pixels at the 
focal plane of the telescope, individual frames were interpolated to a 1\dasec  
grid and the resulting pixel fluxes were summed and averaged.
The photometric stability of 2MASS has been determined to be $\pm 0.02^m$ (Nikolaev et al. 2000). The majority of the images for our survey field were obtained at 
air-masses between 1.2-1.7 and elevation angles of 40-60$\ddeg$.

We converted from magnitudes given in the 2MASS  J, H and K$_s$ bands to 
brightness units in terms of \nwm2sr by using the zero-magnitude J, H and 
K$_s$-band 
factors from Cohen (2001, private communication) of $F_0$ = 1592, 1024 and 667 
Jy with an uncertainty of $\pm 1\%$.  We also note 
that, based on the preliminary absolute calibration constants for the 2MASS, J, 
H and K$_s$ bands, the conversion to AB magnitudes becomes,  $m(AB) =  m_J 
+0.90,  m_H +1.37$ and $m_{K_s} + 1.84$ respectively.
 
Variations in the sky background conditions spanning the five months of data 
obtained for the P565-C field can be identified from a histogram of the 
background pixels in each calibrated field. The shape of this peak was fit by a 
simple Gaussian to determine the mean background level for the frame, and the 
dispersion of the background emission. This mean level was then subtracted from 
each pixel to obtain an image with a zero mean background intensity so that it 
could be co-added with the other frames from this sky region. The mean 
level-subtracted frames were co-added onto a grid with 1\dasec  pixels using the 
individual frame center coordinates as a guide to registering the images. In 
Figure 1 we show the histogram of the fitted dispersion for each of the
images used in this analysis. The data quality as measured by the dispersion of
the Gaussian background fit for each of the 2080 calibrated images shows that
for J and K$_s$ bands, the typical dispersion is $\approx$ 30-70 \nwm2sr 
for more than 90\% of the frames. The H-band data, located near the peak of the 
OH air-glow atmospheric emission, was predictably noisier than either the J or
K$_s$ data with a much broader dispersion of 100-150 \nwm2sr . Although the mean 
intensity of the OH-glow is significant, we found that during the
course of the co-addition process, the noise in this component
integrates down with $N$ images co-added as $N^{-1/2}$.

The 2080 calibrated images were co-added into a final `deep-integration' 
image by using the coordinate information for the center of each image, and 
performing a pixel-by-pixel summation of the intensities. Each image was 
corrected for its own calibrated zero point provided in the image header field  to place the photometry on a common scale.
We also counted the number of measurements that were used in a given pixel to 
establish a statistical weighting function for the summed and averaged data 
which was used in subsequent stages of the analysis. Figure 2 gives the 
histogram of the number of images  available for each pixel, while Figure 3 
shows the final averaged map. The typical number of samples per pixel is 
$\approx$ 500 for an equivalent integration time of $\approx 3900$ seconds. The 
top three images show the J, H and Ks data for a single 7.8-second calibration 
frame. The lower three images are the corresponding deep-integration fields.

The final co-added field was $536\dasec \times 3567 \dasec$ in size with 1\dasec 
pixels. For purposes of Fourier analysis and the computation of the angular 
power spectrum, the field was cropped into seven $512 \times 512$ pixel patches. 
This cropping also removed most of the edge-effects caused by the mosaicking and 
co-adding process. For example, because each calibration scan is offset 5 \dasec 
in Right
Ascension from the previous scan, pixels within 30 \dasec of the east and west 
edges
of the final image are observed significantly less often than pixels within the 
bulk of the
image. The patches are numbered 1--7 and are shown in Figure 4.

Photometry of a random selection of 180 stars in Patch 3 (shown in Figure 3) of the 
field yields the [J-H]:  [H-K$_s$] color-color diagram indicated by the diamond symbol in Figure 5(a). The brighter stars in this field had K$_s = +11.0^m$ 
(S/N $\approx$ 286). Near the 2MASS point source limit of K$_s = +14.5^m$, stars were detected at a typical S/N of $\approx$ 70.
 Also shown in the figure are the colors for  a sample of  
objects in the 2MASS Point Source Catalog at a Galactic 
latitude of $b^{II} = 78^\circ$   
with $K_s \gt +14^m$, and $S/N \gt 10$. The elliptical ring 
indicates the extent of the distribution for $\approx 10^5$ stars. The majority 
of the stars in Patch 3 were detected at low S/N. The photometered stars in Patch 3 appear to have colors 
similar to the ones in the larger 2MASS survey. This finding is also borne out in 
Figure 5(b) which shows the [H-K$_s$] color compared to K$_s$ for the faintest point sources 
within our sample.

\subsection{Sensitivity Limits}

According to the 2MASS Explanatory Supplement, a typical 7.8-second Atlas image 
achieves a 10-sigma detection limit of J = $15.8^m$ (0.7 mJy), H = $15.1^m$ (0.9 
mJy), and K$_s$ = $14.5^m$ (1.2 mJy). The photometry of the faintest stars in 
our deep-integration fields shown in Table 1, along with their S/N values in 
parenthesis, allows us to deduce limits nearly $4.0^m$ fainter in each band
than the standard 2MASS sky survey images. 

By extrapolation from the 2MASS survey photometry, in K$_s$ band, at a detection limit of $10\sigma$, the 2MASS point source 
magnitude limits for the final images will  be $ +17.7^m$ for a 
background noise that integrates down as $t^{-1/2}$. In J and 
H band, these 
limits are $+19.2^m$ and $+18.5^m$ respectively. Generally, the deep-field point 
source detection limits are consistent with those expected from the original 
7.8-second Atlas Images when allowance is made for the increased integration 
times alone. The equivalent AB magnitude limit corresponding to the actual 
$10\sigma$ J, H and Ks band deep-integration limits of $+19.6^m , +19.7^m$ and 
$+17.7^m$ are $ m(AB) = +20.5^m, 21.1^m$ and $+19.5^m$ respectively.

\subsection{Point Source Removal}

With 2MASS and its arc-second-scale resolution, the images of bright stars and 
nearby galaxies can be readily separated from their neighbors
and removed. The seven patches in our 2MASS fields were de-sourced independently 
using the following algorithm: 

We computed the dispersion $\sigma$ of the pixel intensities for each of the 
seven 
patches. Then pixels with 
flux exceeding a fixed number of standard deviations ($N_{\rm cut}$) were 
removed, 
along with their 8 nearest neighbors. The process was 
repeated for all pixels within the image, then the dispersion was recalculated 
to produce the $\sigma$ for the next iteration. This iterative algorithm was 
repeated until no further pixels were found. Typically, only 3-5 iterations were 
needed for the process to converge to a final de-sourced map. We experimented 
with 
$N_{\rm cut}=3,4,5$ and they all gave practically identical results. The results 
below will be presented for $N_{\rm cut}=3$. 

 The magnitude limits of our iterative clipping 
process can be obtained for a given $\sigma$ in \nwm2sr by converting the 
pixel surface brightness limit into an equivalent magnitude per square 
arc-second for a perfectly unresolved star located at the center of the pixel. 
For $\sigma$ = 50 \nwm2sr in J-band, this  corresponds to an integrated 
brightness of a $+23.8^m$ star within a one square arc-second solid angle. 
Similar 
calculations yield $+23.0^m$ and $+22.2^m$ in H and K$_s$ bands. The 
corresponding $m(AB) = +24.7^m, +24.4^m $ and $+24.0^m$ respectively. The 
measured peak surface brightness of an unresolved source seen by 2MASS typically
accounts for 10 \% of the detected emission. This means that actual point source 
magnitude limits for the clipping process are 2.5$^m$ brighter than the 
magnitude limit calculated from the peak surface brightness and the 
clipping parameter.

Table 2 shows the result of applying the clipping procedure to P565-C in the 
K$_s$ band and for all seven patches. Column 1 identifies the patch number 
oriented from south (1) to north(7) along the calibration field. Column 2 gives 
the percentage of pixels that survived the clipping operation. Column 3 gives 
the average number of 7.8-second observations per pixel. In the next three 
columns we indicate the surface brightnesses that correspond to  the clipping 
level at $\sigma$ which is
given in \nwm2sr in parenthesis. (e.g. 36 \nwm2sr corresponds to a
K$_s$ surface brightness of $+20.1^m/$asec$^2$ and a point source magnitude of
$+19.7^m$). For a clipping threshold of $3 \sigma$, all point sources $
1.2^m$ brighter than the magnitudes shown in Table 3 were removed.
Typically, more than 90\% of the pixels in each patch survived the point source
clipping process with 3$\times$3 pixel masking. Given the 7.8-second integration 
per sample, we see in column 3 that the typical integration times for the 
patches
were in the range from 3,682 - 4,056 seconds. 

Figure 4 shows the resultant images  in the three 2MASS bands. The 
seven clipped patches ordered from bottom to top are shown in the left-side 
panels in each band. The masked pixels eliminated by the clipping
algorithm are shown in black. 
The right-side panels in each band show the maps of the number of co-adds that 
went into each pixel.

\section{Analysis}

If the CIB is generated by clustered, luminous  matter in the distant universe,
the structure of the CIB should contribute to the  angular correlation function 
of the background noise fluctuations of the diffuse light in the J, H and K$_s$ 
band data following the removal of the point sources. An alternative, and 
faster, method of characterizing the angular intensity distribution is to 
determine the power spectrum of the 2MASS data, which is the Fourier transform 
of the angular correlation function.

The fluctuation in the CIB
surface brightness can be defined as
$\delta F(\mbox{\boldmath$x$})= F(\mbox{\boldmath$x$}) -
\langle F \rangle$, where $F = \lambda I_\lambda$, $\mbox{\boldmath$x$}$ is the 
two dimensional
coordinate on the sky and $\langle F \rangle$ is the ensemble
average. The two-dimensional Fourier transform is $\delta
F(\mbox{\boldmath$\theta$})= (2\pi)^{-2} \int \delta F_q
\exp(-i\mbox{\boldmath$q$}\cdot \mbox{\boldmath$\theta$})
d^2\mbox{\boldmath$q$}$.
If the fluctuation field, $\delta F(\mbox{\boldmath$x$})$ is a random variable, 
then it
can be described by the moments of its probability distribution
function. The first non-trivial moment is the  {projected 2-dimensional}
correlation function $C(\theta) = \langle \delta
F(\mbox{\boldmath$x$}+\theta) \delta F(\mbox{\boldmath$x$})\rangle$. The
2-dimensional power spectrum is $P_2(q) \equiv \langle |\delta
F_q|^2\rangle$, where the average is performed over all phases. The
correlation function and the power spectrum are a pair of 2-dimensional
Fourier transforms and for an isotropically distributed signal are related
by
\begin{equation}
C(\theta)= \frac{1}{2\pi} \int_0^\infty P(q) J_0(q\theta) q dq,
\label{e1}
\end{equation}
\begin{equation}
P(q)= 2\pi \int_0^\infty C(\theta) J_0(q\theta) \theta d\theta,
\label{e2}
\end{equation}
where $J_n(x)$ is the $n$-th order cylindrical Bessel function. If the
phases are random, then the distribution of the brightness is Gaussian
and the correlation function (or its Fourier transform the power
spectrum) uniquely describes its statistics. In  measurements with a
finite beam, the intrinsic power spectrum is
multiplied by the window function $W$ of the instrument. Conversely, for the
known beam window function, the power spectrum can be de-convolved by dividing
the measured power spectrum by the beam window function.

After Fourier transforming the clipped maps, we computed
the power spectrum for each of the seven patches by 
averaging the square of the Fourier amplitude in concentric rings around a
given wavenumber $q$. The amplitudes were weighted by the number of measurements available for each pixel in the co-added, 2MASS field.
In Figure 6 we show the power spectrum of the data for Patch 3 with
point sources clipped at $3\sigma$, and for an
equivalent integration of 7.8 seconds. The expected 
small-angular scale behavior is dominated by the beam point-spread function.
The large angular scale power spectrum is essentially flat, consistent with
 white-noise distributed fluctuations. At this very short exposure, the 
sensitivity of 2MASS is not sufficient to register large-scale zodiacal or 
cosmological structure, and only the
apparently random and uncorrelated structure of the instrument+atmosphere
background emission are expected to contribute. 
The sensitivity in a single 7.8-second image is, however, adequate 
to detect stars and galaxies at small
angular scales, where we see a power spectrum that resembles white noise
 multiplied by the instrument beam profile. The beam window function can thus be 
determined very accurately and the power spectrum de-convolved. The de-convolved 
power spectrum is shown with plus signs.
 
Figure 7 shows the de-convolved power spectrum of the deep-integration clipped 
Patch 3 in K$_s$ band with Poissonian error bars (i.e. $\delta P/P = $[number 
of points within $dq$ of $q$]$^{-1/2}$). The ``x" signs correspond to the 
maximal 
exposure clipped images. The filled diamonds correspond to the atmospheric 
contribution to the power spectrum (i.e. the power spectrum from one co-add 
divided by the mean number of co-adds for this patch). One can clearly see the 
excess power, and the different slope of the power spectrum, for the diffuse 
emission in the co-added image. This suggests a significant contribution from 
background infrared emission below the clipping threshold which includes
the CIB emission. Similar results obtain for the 2MASS J and H bands.

As Figures 3 and 4 show, there are residual stripes in the long-exposure images. 
Such stripes arise from variety of known instrumental effects including 
diffraction
within the telescope optics, and electronic crosstalk in the presence of  
bright stars. 
In Fourier space, the horizontal and vertical stripes collapse onto 
the wavenumber axis, corresponding to zero frequency in either horizontal or 
vertical direction. 
Figure 8 shows the Fourier map of the maximal exposure Patch 3 
in K$_s$ band. Similar maps obtain for J and H bands and are not shown for 
brevity.
The Fourier 
transform of the stripes 
is clearly seen in the cross-pattern along the axes. In order to de-stripe the 
images and compute the underlying power spectrum for the diffuse emission, we 
excised the narrow strip of pixels along each axis of the Fourier transform. 
The final power spectra are not 
sensitive to the width of the removed strip, and in what follows we present the 
numbers with 2 pixels removed in the Fourier plane along each axis. 

Figure 9  presents the variation of the background fluctuation 
dispersion in each of the 
seven clipped patches before and 
after de-striping for the J (triangles), H (squares) and K$_s$ (crosses) 
bands. The de-striped and ``raw" dispersions are proportional, 
which means that the removed variance due to stripes 
does not come from the residual sky emission.

Figure 10  shows the final power spectrum of the diffuse 
emission in the seven 
patches of the field after de-striping. The expected atmospheric power spectrum 
for the final co-add would be given by the power spectrum at 7.8 sec (Figure 
6) divided by the number of co-adds in the final maps and shown with 
filled diamonds. The final rms fluctuation in 
the surface brightness for the patch after de-striping is is shown in the upper 
left corner of each patch in 
units of \nwm2sr .
The persistence of a distinct power law and its similarity in each band shows 
that 
it is a robust signal
reflecting a common background structure in this direction of the sky.  

Note that our method subtracts out the mean (DC) level of the background, including the CIB. The latter therefore cannot be constrained directly by the fluctuations method. However, the DC levels of the near-IR CIB have been measured directly from the COBE/DIRBE and IRTS data. E.g. they give ca. 50 \nwm2sr  in J band (Cambresy et al 2001), ca. 25 \nwm2sr in K (Wright \& Reese 2000) and are consistent with the IRTS measurements of the CIB mean levels between these bands (Matsumoto et al 2002). We clip galaxies out to K$\sim 18.5^m-19^m$ so the clipped galaxies contribute only $\lsim 6$ \nwm2sr and the K-band galaxies observed down to the faintest magnitude of $K\sim 24^m$ contribute $\sim 9$ \nwm2sr (see Fig. 2 of KOMSC). In principle, the mean level of the CIB can be constrained from the TEV gamma-ray measurements of galaxies with known energy spectrum (Stecker \& De Jager 1993), but these constrain the CIB mean levels most effectively longwards of 5 $\mu$m, where no direct measurements of the CIB mean levels currently exist.

\section{Non-Cosmological Contributions}

It is clear from Figure 6 that the final images contain a diffuse infrared 
component with a power spectrum that is distinct from the white noise 
contributed by the atmosphere+instrument background and by stars. 
We now investigate some of 
the potential sources of this signal, and assess their contributions.

\subsection{Atmospheric Effects}

The largest contribution to the sky background comes from the atmosphere 
itself, which for a single 7.8-second integration at K$_s$ band amounts to 295  
MJy/sr or equivalently +12.4$^m$/arcsec$^2$. The spectrum of the emission
has the color shown in Figure 4 represented by the square plotting symbol and is dominated by a combination of atmospheric and instrumental thermal emission.
The 2MASS frames preserve the observed background sky levels measured relative 
to camera dark frames with the shutter closed. This background is normally 
largest in the K$_s$ band, although it can be even larger in the H-band due to 
atmospheric OH air-glow emission. The only background compensation made during 
Atlas Image construction is to adjust the six survey 
frame backgrounds by a constant to 
produce seamless co-added Atlas images with 7.8-second integration times. 
Because the OH air-glow (especially at H-band) 
contains structure on scales at, or below, the 2MASS frame size, the 
resulting 2MASS Atlas images often show large background variations. 
According to Ramsay et al. (1992) the variation of OH-glow due to 
gravity waves occur at angular scales of about 17 degrees and with time scales 
of 11 minutes. 2MASS analysis of short-term, background noise characteristics
in the Atlas images suggests that much finer spatial scale variations also 
exist. 

Since variations of the mean atmospheric background level contribute to
increasing the final background noise of the co-added image, we removed this 
component in the individual images via a simple Gaussian fit, which determined the amplitude and dispersion ( $\sigma$ ) of the background pixel intensities in each image. This Gaussian fit was truncated at the high end to eliminate the contributions from stars. We then subtracted the Gaussian-fitted, 
mean background intensity  from each pixel. Thus all information about the 
mean intensity of the background has been removed from the final co-added data, 
including any information about the mean intensity of the various astronomical
backgrounds. We can, however, evaluate the contributions by some of these 
backgrounds to the `noise' of the final co-added image.
 
The image dispersions determined from the individual Gaussian fits were compared with airmass in Figure 11 to 
determine whether any obvious trends existed in the data that could suggest a 
correlated atmospheric 'signal' spanning the five months of data. No apparent 
correlations were identifiable. Given a mean sky background noise near $\approx 300 $ 
MJy/sr in K$_s$ band, Figure 11 shows that over the course of 5 months, the 
background noise was surprisingly uniform 
(typically $\pm 500$ \nwm2sr ), and varied by less than $\pm 100$ \nwm2sr .
 In
H-band where OH-glow dominates, the variation of the background 
noise was $\pm 1000$ \nwm2sr for a mean sky intensity of 140 MJy/sr . 

Figure 12 shows the histogram of the residual pixel intensities, F, in 
Patch 3 after clipping, where we define the  probability density as the 
ratio of the number of pixels in the $[F;F+dF]$ range of brightness to the total 
number of pixels in the patch. Although the individual images that make up the 
co-added image have been corrected to a zero mean computed in the absence of stars, the histograms  
show the effect of the variation of image-to-image 
pixel noise for the co-added sample. 
The histograms have offsets of 190, 250 and 330 
\nwm2sr for K$_s$, J and H-bands respectively.
This corresponds to the dispersion of the individual
frame background noise values shown in Figure 11 determined in a 
consistent manner for the pixels in Patch 3.
It is also evident that the histograms are smooth and Gaussian-like with some 
evidence for tails toward larger values resulting from the remaining un-clipped 
point source emission. This indicates that transient 
atmospheric conditions leading to increases in the individual image noise are
sufficiently Gaussian for our purposes, with no long-term correlation over the
5-months of data. 

Figure 5(a) shows that, based on the color of the atmospheric component (box symbol with 1-$\sigma$ error bars) determined from the 2080 individual images 
of [J-H] = +1.9$^m \pm 0.4^m$ and [H-K$_s$] = +1.2$^m \pm 0.4^m$ it is also evident from the
colors of the residual emission in Patch 3 and plotted with an asterisk (e.g. [J-H] = 
$+0.07^m$ and [H - K$_s$] = $+0.57^m$) that no trace of an 
atmospheric component 
remains in the final data. This confirms that by removing the mean background 
from each image prior to co-adding, we have eliminated the majority of the 
atmospheric contamination. The mean residual background intensity in K$_s$ band 
was 0.14 MJy/sr or equivalently  190 \nwm2sr and $+20.8^m$/asec$^2$, which
is less than  0.05\% of the original sky background 
detected in a single 7.8-second image. This confirms the long-term 
stability of 
the atmosphere and shows that its properties can largely be assumed to be 
uncorrelated in both space and time over the angular scales we are investigating
in this study. 

\subsection{Instrumental Effects}

The  data that we use span over 5 months of
observations. As a consequence of the observing protocol, the telescope
was moved $5\dasec$ in RA during each N-S scan, and the mosaicking process 
insured
that, in the regions of overlap, multiple array pixels were used to measure
the brightness of a particular line of sight.
The more than 500 brightness measurements made in a typical pixel
of the final deep-integration image are the product of measurements made by 
$\approx$ 20 different pixels in the array itself, thereby
reducing greatly the possibility of systematic gain and responsivity effects
entering the final image.

We can characterize the short-term instrument and background changes which
lead to non-zero residual emission in the co-added images by simply differencing
the `odd' and `even' images in the 2MASS data on a pixel-by-pixel basis. We 
refer 
to these difference maps as A--B maps. 
Observationally, we have for every pixel, a time-ordered string of
measurements that were averaged together to obtain the pixel value in the
deep-integration image. We can separate this string of measurements, 
one per image, into an even and an odd series, and subtract the individual pairs
of measurements. In doing so, we remove any time-independent component 
which appears as a mean `DC' level. The new image based on these differences is 
now a repository for any background component that varied over the course of 
consecutive nightly measurements of the same pixel. For example, if the 
background conditions during a single night varied significantly due to changes 
in OH-glow between the consecutive observations of the calibration field,
a non-zero A--B residual will occur for some or all the pixels
in the calibration strip. If this difference persists over the course of 
many nights, then the long-term average of the A--B over the full data will
produce a statistically significant residual. Instrumental effects, such as 
long-term responsivity changes and drifts will also be included if the  
conditions changed during the course of consecutive nights of observation.

Figure 13 shows the resulting A--B co-added image for Patch 3 in K$_s$ band, 
masked according to the clipped output to remove regions containing emission 
from point sources. The residual emission in the un-blanked portions correspond 
to $\sim$ 2 \nwm2sr for the 2MASS bands. 
In Figure 14, the power spectrum of the A-B image confirms that there is 
essentially no correlated signal in the background data that might be 
contributed by instrumental variations
or intra-day variations in atmospheric emission. The power spectrum is white 
noise 
modulated on small scales by the window function due to IPAC interpolation to 
the 
1\dasec pixels.

Latent star-like images can be created by the NICMOS detector material after 
exposure to bright stars. The decay time constant is about 10 seconds. This 
process produces a string of point sources of decreasing brightness that are 
spaced with offsets that match the 83.3$\dasec$ sequential steps used in creating the 
overlapping frames at the telescope. After final processing of the deep-integration images, we see no signature of this feature in the power spectrum.

Internal telescope reflections can also produce stripe/streak features
extending over $ 1\ddeg$ in radius from the bright star's center. Internal 
reflections of bright stars or other starfield-dependent instrument artifacts 
cannot account for the detected fluctuation power spectrum.
As we noted in Section 3.1, there is no evidence that the variation of unclipped 
artifacts in the seven patches (e.g. Patch 2,6,7 vs. Patch 3,4,5) leads to a 
significant change in the power-law feature seen persistently across
all the patches and bands, at least over angular scales from 2\dasec to 
100\dasec . 

As a consequence of a variety of transient events such as cosmic ray hits, 
and meteor streaks, the 2MASS images are automatically blanked by the
2MASS pipeline software wherever 
such `bad pixels' appear. This means that the number of samples per pixel 
in the initial Atlas images may
be less than the canonical value of 6 for some small percentage 
of the pixels in the field. This impacts the actual effective integration times 
used in calculating our final co-added, 
background noise averages across the field. Since fewer than
1 \% of the pixels are typically involved per 1.3-second exposure, and since
there is no correlation of these events across the full 5-month archive 
of images, these pixels are randomly distributed across the image and 
do not contribute 
to a systematic `pattern noise'. This is also apparent in the white 
noise-character of the co-added image power spectrum at scales of a few 
arcseconds where pixel blanking would normally contribute.

The difference maps over the area remaining after source clipping give flux 
dispersions a factor $>10$ lower than that of the co-added clipped images. E.g 
for patch 3 we get $\sigma=(2,3,2)$ \nwm2sr in J, H, K$_s$ bands respectively. 
This 
adds in quadrature to the variance from other terms including the CIB. Thus the 
instrument noise is negligible compared to the signal 
in Figure 10.

\subsection{Stray Light}

We considered the possibility that stray light and stray electronic 
signals in the telescope and instrument could mimic large scale 
fluctuations.  Even the ideal Fraunhofer diffraction point source 
response function from a telescope falls off with angle only as 
$~1/\theta^3$, and the integral of the light outside $\theta$ falls 
off only as $1/\theta$.  Diffraction from dust grains and other 
imperfections on the optical elements, and in the Earth's atmosphere, 
may be expected to have similarly slow asymptotic behavior.  Hence an 
experimental test is required.

We therefore found 2MASS observations of several bright star fields near the Galactic plane where the star density after 7.8 sec exposure is comparable to our Patch 3. The fields, 1 degree in radius, were  located at the J2000 coordinates (RA , Ded ) = ($19^h$ 18$^m$ 23.6$^s$, -37$^\circ$ 5$^\prime$ 30$^{\prime\prime}$) and ($18^h$ 6$^m$ 32.0$^s$, -39$^\circ$ 31$^\prime$ 19.7$^{\prime\prime}$). The short exposure images of these regions look 
qualitatively much like the long exposure co-adds in darker regions, 
and we expect that the stray light and stray electronic signals 
should also resemble those in our deep regions when properly scaled. 
We analyzed these data with the same clipping algorithms and computed 
the residuals.  The clipping algorithm leaves behind about the same 
number of pixels (94\%) as for our deep exposures.  However, we found 
that the clipped power spectra are almost perfectly flat, except for 
the very largest spatial scales.  This clear qualitative difference 
shows that stray light does not explain the slopes of our measured 
fluctuation power spectra.   The clipping algorithm leaves power 
spectrum residuals that are 2.4 to 3.3 orders of magnitude lower than 
the unclipped power spectra, showing that at least 93\% of the total 
flux of each star is removed by the algorithm.

Figure 15 shows these results, comparing the unclipped and clipped 
power spectra of our one such long exposure region (Patch 5) with the results of the 
bright star field from the main 2MASS survey, scaled to match the unclipped power spectra. 
The scaled clipped bright star field power spectrum is significantly 
fainter than the long exposure region, showing that stray light from 
the stars is unlikely to cause the measured fluctuations on large 
angular scales.

All the patches on close visual inspection showed residual extended halos around the brightest stars. To test whether extended halos around stars cause a significant effect on the power spectra, we analyzed Patch 3 with SExtract, determined the list of sources and their isophotal diameters, and made a new map with square masks with edges that were four times the isophotal diameters.  We then computed the power spectra again. The result was within 2\% for scales below 1 \dasec, with a decrease of 10\% at 10 \dasec . While these changes are within the error bars, they do show a small trend, and would change the power spectrum slope by about 0.04.

\subsection{Zodiacal Emission}

At wavelengths from 1.25 to 2.2 \um the emission from interplanetary dust
is dominated by scattered sunlight which declines in brightness by 3-fold
over this wavelength range. 
DIRBE photometry in the direction of the 2MASS field obtained near a solar 
elongation angle of 90\ddeg yields an integrated K-band 
background of
$\sim 150$ \nwm2sr with a spectrum dominated by scattered Zodiacal light (ZL). 
Moreover, the color
of the ZL follows the solar spectrum to $\approx$ 2 \um .  In 
Figure 5(a), the color of ZL is also indicated (large triangle symbol) and it is 
clear that the residual background seen in the 2MASS co-added image is 
significantly redder. This indicates that any residual
 sky contribution by ZL to the final co-added image appears to have
 been subtracted out during the co-adding
process itself. Indeed, because we are only concerned with the detection of
structure in the CIB, some amount of ZL can be tolerated in
our analysis provided that this emission is smooth over the angular
scales of interest. In our 
analysis, we are not concerned with the absolute magnitude of the ZL
but rather with its angular structure at 2MASS
scales from  2\dasec - 100\dasec . It is generally believed that ZL
is smooth at these scales due to a variety of physical mechanisms
(e.g. Poynting-Robertson drag, orbital phase diffusion) which act
to smooth out dust irregularities which would appear at these scales.

Cometary dust bands are known to exist at $\beta < |10^\circ|$ with half 
width of $\approx 3^\circ$ (Reach, 1992). These features are significantly
below the latitude of the 2MASS field and are not relevant to this study. Some 
claims have  been made for fainter bands at $\beta  = \pm 22^\circ$ 
(Sykes, 1988) but these have not been confirmed in other studies. In all cases, 
the banded features are degree-scale features, 
prominent in the mid-IR near 25 \um and are not known to 
contribute to scattered light emission in the 1-2 \um range. A resonant 
dust ring outside Earth's orbit was detected by DIRBE at 4.9 - 25 \um 
(Reach et al. 1995), however, its emission peaks at 25 \um ,  and at 4.9 \um its
excess surface brightness above the local Zodiacal background
is $\approx 0.05$ MJy/sr (30 \nwm2sr ). This feature is
located at an elongation angle of $90^\circ$ and extends $15^\circ$ 
in longitude with a $30^\circ$ FWHM in latitude. This places the 2MASS region
outside the zone where the emission from this feature predominates.

Studies of the arcminute-scale angular structure of the ZL by 
Abraham, Leinert and Lemke (1998) with the ISO/ISOPHOT instrument also
indicate that at 25 \um , no structure can be detected above 0.2\% of the 
total ZL brightness in a number of fields distributed 
over a range of 
Ecliptic latitudes.  Because the thermal emission is optically thin and the
dust grains are essentially isothermal, the distribution of the scattered
light should follow the variations in the thermal emission. 
Based on a total
 2.2 \um surface brightness for the ZL of 130 \nwm2sr 
from DIRBE photometry,
and the limit set by the structure at 25 \um , we expect that ZL 
structure
will not contribute more than $\approx 1$ \nwm2sr in the 2MASS bands
on arc-minute scales. 

The 2MASS observations which were used in this study, span a time period from
April to August and a solar elongation range from  $115^\circ$ to
$128^\circ$, which means that any persistent Zodiacal structure would become
uncorrelated during the data averaging process. 
We, therefore, conclude it is unlikely that sub-arcminute structure in the 
Zodiacal background emission is present in the 2MASS data in the form of
a well-defined power spectrum.

\subsection{Interstellar Dust}

Although interstellar cirrus is primarily a feature of the far-infrared
sky, the presence of PAH emission in cirrus material allows some interstellar
clouds to  contribute to the infrared backgrounds even at near-IR wavelengths. 
Studies of PAH emission show a spectral feature at 3.3 \um and no other 
identifiable lines at 1-2 \um in the 2MASS band-passes. This implies that
there is no radiation mechanism capable of producing foreground emission within
the 2MASS bands that could generate infrared background fluctuations. 
Moreover, investigations of the angular structure of cirrus
emission have found structure as small as the
IRAS resolution limit near 2 arcminutes. Gautier et al. (1992), Wright (1998) 
and 
KO find that cirrus follows a power law of the form $P(q) \propto
q^{-n}$ where $n \approx 2.5-3.0$. In addition Guhathakurta \& Cutri (1994) find that the power spectral slopes of the cirrus emission detected in the optical V band down to a few arcsec scale was the same as for the IRAS emission at long wavelengths. This shows that the cirrus structure at small scales and shorter wavelengths, probed by the 2MASS survey, is consistent with the longer-wavelength characterizations.

Interstellar dust can, in principle, cause extinction effects that under some 
circumstances could modulate the light from a more distant and uniform star 
background, giving rise to structure in the infrared background. Because the 
amount of extinction varies strongly with wavelength, the strongest fluctuations 
caused by a clumpy interstellar medium would occur in J-band, and with declining 
strength in H and K$_s$ bands. No such wavelength-dependent variation in the 
power spectra in Figure 10 is apparent. At K$_s$ band, models for the amount of 
interstellar extinction indicate about A = 0.05$^m$ and a differential 
extinction across the $1^\circ$ field of $\approx 0.005^m$. This would
correspond to a modulation of a uniform background source by 0.5\% at the 
$1^\circ$-scale. Smaller-scale extinction variations with a higher amplitude may 
also be possible if the mechanism involved ( e.g. cirrus) has structure at these 
scales. We can estimate the amplitude of the $1^\circ$ component by assuming 
that, at most, the background emission is produced by unresolved stars below the 
clipping threshold. From Table 2, a typical clipping threshold ($3\sigma$) is 
about 150 \nwm2sr so that for a 0.5\% modulation of the extinction
one obtains an amplitude of about 1 \nwm2sr .  

\subsection{Diffuse Galactic Starlight}

As we note from Table 2, the 3-$\sigma$ clipping threshold in Patch 3
eliminated stars brighter than approximately $+18.5^m-19^m$ in K$_s$-band.
The color-color and color-magnitude diagrams in Figures 5(a) and 5(b) show that
the properties of the extracted point sources are consistent with ordinary
stars.  
We can compute the [J-H] and [H-K] colors of the residual noise in the
clipped fields which amounts to 351 $\pm$ 55, 236 $\pm$ 48 and 
196 $\pm$ 34 \nwm2sr for J, H and K$_s$ respectively, which leads to [J-H] =
 $+0.07^m$ and [H-K$_s$] = $+0.57^m$. In
Figure 5(a) we find this color to be
distinct from Zodiacal light emission ([J-H] = +0.4$^m$ and 
[H-K$_s$] = -0.4$^m$), or atmospheric emission ([J-H] = +1.9$^m$] and
[H-K$_s$] = +1.2$^m$), but 
similar to that of the population of point sources that were 
photometered. In fact, it is similar to the color of K0V stars 
([J-K] = +0.64$^m$).

Based on the photometry of the point sources in Patch 3, we can determine the
star count distribution for this region and compare it with star count 
models to assess
how much starlight remains in the field after bright-source clipping. 
We have calculated the expected star counts
in the direction of the 2MASS field by employing 
the Faint Source Model (FSM) developed by Arendt et al. (1998) for the analysis 
of the COBE/DIRBE data. The FSM is based on the five-component galaxy population
model described by Wainscoat et al. (1992). Given the galactic coordinates of the 
field, only two components contribute to the star counts:  A halo population and 
a disk population. The FSM model consists of 87 types of stars spanning a 
luminosity range $ -6.0  \geq M_v \geq  +8.5 $, and assigned to each of the 
structural 
components as appropriate. An integral is performed along the line-of-sight to 
the specified region, and at each distance increment, a calculation 
of the number of stars and their apparent magnitudes is performed. The results 
are stored in an array that yields the cumulative number of stars brighter than
magnitude m, $N(m)$ and is presented in Figure 16. We also show in the same 
figure the actual star counts obtained in 6 magnitude bins within Patch 3. Due 
to the small number of stars brighter than +16$^m$ in this $8^\prime \times 
8^\prime$ field, the  statistical error bars for the brightest two bins are 
significantly larger than for the fainter bins plotted. Generally, the actual 
star counts follow the predicted counts quite closely over the range from +16$^m 
$ to +20$^m$. We note, however, that the most significant counts between +17$^m$ 
to +19$^m$ are higher than the predicted counts, reflecting the fact that faint 
galaxies should enter the statistics for these fainter counts. 

Because Galactic stars have a white noise spatial distribution, the flux
fluctuation, $\sigma_*$ from stars brighter than $m_{\rm cut} $ is given by: 
$\sigma_*^2 = \omega_{\rm beam}^{-1} \int_{m_{\rm cut}}^\infty I_\nu^2(m) n(m)
dm$ where $\omega_{\rm beam}$ is the 2MASS beam area, $n(m) dm$ is the number of
stars within $dm$ of the apparent magnitude $m$ and $I_\nu(m) = I_{\nu,0}
10^{-0.4 m}$ is the intensity of a star of magnitude $m$ at frequency band
$\nu$. Using a Galactic model normalized to 2MASS, we estimate that stars 
fainter
than $m_{\rm cut}\geq 18.5$ would contribute less than $\sigma_* < 9$\nwm2sr to
the total fluctuations. This component adds in quadrature to the total
dispersion, thus contributing less than 2-3 \% of the total signal in the 
co-added 
maps.

The power spectrum of an unclipped sky patch, which is dominated by the
emission from foreground stars, shows that stars are not clustered, but
have a white-noise spectrum for angular scales 1\dasec - 100\dasec . 
Our study of the properties of the remaining point sources near, or just below
the $3-\sigma$ clipping threshold indicates colors that appear to be similar to those just above
this threshold, and we conclude that there is no firm evidence that the unresolved stars
contributing to the diffuse emission in the field are a different 
population than the bright stars that were removed, or that they 
have different clustering properties. Although the diffuse stellar component
dominates the background light, it is not structured in a way that is expected 
to lead to a power-law spectrum. 

We checked whether the imperfect blanking of starlight in the field
may have had an impact upon the resulting power spectrum. We examined this in
several ways : 1) we ran the source extractor from the COBE DIRBE fluctuations 
work (see KO for discussion), 2) employed the widely-used package SExtractor 
(Bertin and Arnouts 1996) to identify discrete
sources in the 2MASS fields, and 3) simulated the effect of the residual halos 
on the large scale power spectrum of the diffuse emission. 
The SExtractor algorithm produces a file which specifies the integrated 
magnitude, location and shape of all extracted sources brighter than a 
user-specified S/N. We used this information to create a mask of all the 
discrete sources, and this was then used to blank out the sources. 
SExtract identified 581 discrete sources brighter than 3-$\sigma$ in Patch 3 at 
K$_s$ band, which corresponded to a masking of 19,160 pixels or 7.5\% of the
total field area of Patch 3. This is identical to the level of masking obtained
by our clipping algorithm. Figure 17(a) shows that the angular radii 
of the sources determined by SExtract are $\lt 2\dasec$ and therefore unresolvable by 2MASS. 
Figure 17(b) also shows the variation of the rms source radius with integrated
K$_s$ magnitude determined by SExtract, which indicates that the fainter 
sources in this study are substantially unresolved.
A comparison 
of the first two procedures showed agreement between the sources that were 
blanked. 
SExtract was more efficient in removing the extended faint halos of the
brighter stars from the sky data, however, there was no practical effect 
on the power spectrum of the diffuse 
emission for the remaining field pixels. Nevertheless, we checked by 
independent 
simulations that the remaining halos do not mimic the observed signal. The 
halos retain the white noise distribution of the underlying Galactic stars and 
would produce the white power spectrum on scales outside the 2MASS beam ($\sim 
2\dasec$) contrary to what is measured in Figure 10.

\subsection{Nearby Galaxies}

The nominally complete 2MASS sensitivity limit at 10-$\sigma$ to galaxies as point sources is K$_s = +14.5^m$ (or +16$^m$ at 3-$\sigma$). Studies by Huang et al (1997) and Cowie et al. (1996)    
with the Hawaii K-band Galaxy Survey and a magnitude limit of $K < +16^m$ 
find a corresponding
redshift limit near z $\approx$ 0.2. 
There are no discernible bright galaxies 
in the P565-C field of view with near-IR
diameters significantly larger than $\approx 5\dasec$, and magnitudes 
brighter than $\approx +16^m$ which indicates that 
in this sky direction there are no large ($r \approx $20 kpc)
galaxies closer than $\approx$ 400 $h^{-1}$Mpc ($\theta$ = 5\dasec ; z $\approx$ 
0.1). At the fainter magnitudes covered by this survey, the situation becomes 
more complicated.

Field spheroidal galaxies with $M \approx 10^{11} M_\odot$ are expected to
correspond to objects with $+18 \lt K \lt +20$ and a redshift range
$1 \lt z \lt 2$ (Kauffmann and Charlot 1998, Jimenez \& Kashlinsky 1999, KOMSC). 
The Hawaii 
Medium Deep Survey covers 170 arcminutes$^2$ to a magnitude limit of K = +19.
Spectroscopic studies of the galaxies in this field by Cowie, Songaila, Hu and 
Cohen (1996)
show an approximate correlation between K-band magnitude and redshift. 
Our faintest (S/N=$3 \sigma$) unresolved sources 
near K$_s \approx 19.5^m$ would include galaxies
with redshifts in the range $ 0.4 \lt z \lt 1.5$ if they were present in the 
field.
Deep K-band galaxy counts by Gardner (1995) yield an expected surface 
density of
galaxies with magnitudes between $+18^m$ to $+20^m$ of 10,000/deg$^2$, which 
implies that in Patch 3 we should see $\sim 200$ galaxies. 
The contribution by faint galaxies to the integral star counts for Patch 3 
is shown in Figure 16. The figure shows that, at the galactic latitude of Patch 3, the contribution to the point source counts by distant, faint galaxies 
is significantly smaller than the expected number of foreground stars at the
same latitude. Discrete source counts are, therefore, dominated by foreground 
stars for magnitudes brighter than $\approx +20.0^m$.
Figure 17(b) also shows, for the sources identified by SExtract, that the 
discrete sources in Patch 3 fainter than +17$^m$ in K$_s$ band are completely 
unresolved by 2MASS.
We conclude that our clipping algorithm removes a significant number, and 
very likely all, of the resolvable galaxies  to redshifts of $z \approx 1$, 
although our clipping procedure may still
leave dwarf galaxies, or other local, 
low-surface brightness systems present in the data. 

Based upon the previous discussions, we can summarize the likely contributions
to the background fluctuations from non-cosmological foreground
sources in Table 3. In columns 2 and 3 we indicate the estimated color of each 
component. In column 4 we
give the mean background intensity over the full field-of-view, and in column 5
we estimate the background fluctuation expected at a scale of approximately one 
arcminute.

\section{Conclusions}

By using the calibration field data from 2MASS, we have created a 
deep-integration field at J, H and K$_s$ bands covering an area of $8.5^\prime 
\times 1^\circ$. After clipping out all
foreground point sources brighter than the K$_s$ magnitude of $\sim +19^m$ 
 we detect a statistically significant structure of the residual 
diffuse emission. The angular power spectrum of this emission has a power law 
dependence with the slope expected from that of the CIB due to clustered 
galaxies. The cosmological part of this and its implications are discussed in 
the companion paper KOMSC. 

We demonstrated in this paper that the various foreground (atmospheric, zodiacal 
and Galactic) contributions to the residual diffuse background are small and in 
any case would be of different angular slope. The color of these components 
would likewise be very different from the color of the found signal. This is 
consistent with the signal being due to CIB from early galaxies.

We note in conclusion that the amplitude of the signal is broadly consistent 
with other findings on larger angular scales by KO from the COBE DIRBE and 
Matsumoto et al. (2000, 2002) from the IRTS mission. E.g. the KO analysis suggested a CIB 
fluctuation of $5.9^{+1.6}_{-3.7}$ \nwm2sr in the DIRBE K band at $\sim 
0.5^\circ$ scale when all galaxies are included. That was also confirmed in the 
IRTS findings of Matsumoto et al. (2000, 2002). In this study we detect an amplitude in 
K$_s$ band of $\sim 10$ \nwm2sr at $\sim 1.5^\prime$ after removing all galaxies 
out to K$_s \simeq 18.5$ consistent with the above after accounting for the 
difference in angular scales and the contribution from the removed galaxies 
(KOMSC). Our results lend further support to the emerging body of evidence that 
the CIB levels are significantly higher than what can be produced by the 
observed populations of galaxies (KO, Wright \& Reese 2000, Cambresy et al. 
2001).
 
{\bf Acknowledgments} 

This publication makes use of data products from the Two Micron All Sky
Survey,
which is a joint project of the
University of Massachusetts and the Infrared Processing and Analysis
Center/California Institute of Technology,
funded by the National Aeronautics and Space Administration and the National
Science Foundation.

We would like to thank Dr. Rick Arendt for his careful reading of the 
manuscript and his assistance in using the DIRBE Faint Source Model and 
SExtract. We would also like to thank Ms. Destiny Coslett at Long Reach 
High School in Columbia, Maryland for her assistance in the photometry of
the faint sources in Patch 3. Her assistance in identifying and
cataloging standard star
field scans in the 2MASS scan archive was also greatly appreciated.

{\bf REFERENCES}\\

Abraham, P. et al. 1998, in `The Universe as seen by ISO', P. Cox and M.F. Kessler eds., ESA Publication Division, Noordwijk, pp 145-148.\\
Arendt, R. et al. 1998, Ap.J., {\bf 508}, 74.\\
Beckwith, S. et al. 1976, Ap.J., {\bf 208}, 390.\\
Bernstein, R.A., Freedman, W.L. and Madore, B. 2002, Ap.J (In Press)\\
Bertin, E., Arnouts, S. 1996, A\& A, {\bf 117},393.\\
Cambresy, L. et al. 2001, Ap.J. {\bf 555}, 563 \\
Casali, M.M., and Hawarden, T.G. 1992, JCMT-UKIRT Newsletter No. 4, 33\\
Cowie, L., Songaila, A., Hu, E. and Cohen, J. 1996, AJ {\bf 112}, 839.\\
Cutri, et al. 2001. 2MASS Explanatory Supplement.\\
  http://www.ipac.caltech.edu/2mass/releases/second/doc/explsup.html\\
Djorgovski, S. et al. 1995, ApJ (Letters) 438, L13.\\
Dwek, E. and Arendt, R. 1998, Ap.J. {\bf 508}, L9. \\
Dwek,E., et al. 1998, Ap.J., {\bf 508},106.\\
Fall,S.M., Charlot, S. and Pei, Y 1996,Ap.J.Lett,{\bf 464},L43.\\
Gardner, J.P., 1995, Ap.J. {\bf 452}, 538.\\
Gardner, J.P., et al. 1997, Ap.J.,{\bf 480}, L99.\\
Gautier, N. et al., 1992. AJ, {\bf 103}, 1313.\\
Gorjian, V. Wright, E. and Chary, R. 2000, Ap.J. {\bf 536}, 550.\\
Guhathakurta, P. \& Cutri, R.M. 1994, in `The First Symposium on the Infrared Cirrus and Diffuse Interstellar Clouds", ASP Conference Series, Vol. 58, 1994, R.M. Cutri and W.B. Latter, Eds., p.34\\
Hauser, M. and Dwek, E. 2001. Ann. Rev. Astro. and Astrop. {\bf 39}, 249.\\
Huang, J.-S., et al., 1997, ApJ, 476, 12.\\
Kashlinsky, A., Mather, J., Odenwald, S. and Hauser, M. 1996,Ap.J.,{\bf 470},681.
(Paper I)\\
Kashlinsky, A., Mather, J., and Odenwald, S. 1996, Ap.J.Lett.,{\bf 473},L9.
(Paper II)\\
Kashlinsky, A.  and Odenwald, S. 2000,Ap.J.,{\bf 528}, 74. (Paper III)\\
Kashlinsky, A., Odenwald, S., Mather, J., Skrutskie, M. and Cutri, 
R. 2002, Ap.J.(Letters), submitted.  (KOMSC)\\
Kauffmann, G. and Charlot, S. 1998, MNRAS {\bf 297}, L23\\ 
Madau, P. et al. 1996, MNRAS, {\bf 283}, 1388.\\
Malkan, M. and Stecker, T. 2001, Ap.J, {\bf 555}, 641. \\
Matsumoto, T. et al. 2000, in `ISO Survey of a Dusty Universe', D. Lemke, M.\\ 
Matsumoto, T. et al., 2002, Ap. J., (Submitted)\\
Nikolaev, S. et al. 2000, A.J. 120, 3340.\\
Stecker, F. W. \& De Jager, O. C. 1993, Ap.J.,415,L71\\
Stickel, eds. (Berlin and Heidelberg: Springer-Verlag), 96.\\
Persson, S., et al., 1998 AJ, {\bf 116}, 2475. \\
Ramsay, S.K. et al., 1992. MNRAS {\bf 259}, 751. \\
Reach, W. 1992, Ap.J, {\bf  392}, 289. \\
Reach ,W. et al 1995, Nature {\bf 374}, 521.\\
Skyes, M., 1988, Ap.J(Letters) {\bf 334}, L55.\\
Vogeley, M. 1997, AAS. 191, 304. (astro-ph/9711209).\\
Wright, E. L., 1998, Ap.J, {\bf 496}, 1.\\
Wright, E. L. \& Reese, E.D. 2000, Ap.J., 545, 43.\\

\newpage
\vskip 0.2truein
\noindent{Table 1. Selected faint stars in the P565-C region}\\
\begin{tabular}
{l l l l l l l l l}
Star&$m_J$&$m_H$&$m_{K_s}$&[J-H]&[H-K$_s$]\cr
\hline
4&20.0 (8.4)&19.2 (4.4)&18.9 (6.5)&+0.8&+0.3\\
11&19.4 (9.2)&18.5 (16.4)&18.3 (5.0)&+1.1&+0.2\\
16&20.7 (4.5)&20.5 (2.9)&19.4 (3.9)&+1.3&+1.1\\
17&20.0 (7.5)&19.3 (2.4)&19.2 (5.0)&+0.8&+0.1\\
22&19.7 (4.3)&19.7 (2.6)&19.4 (4.2)&+0.3&+0.3\\
28&20.2 (2.6)&19.5 (3.2)&18.6 (4.0)&+1.6&+0.9\\
\hline
\end{tabular}
\vskip 0.5truein

\newpage
\vskip 0.2truein
\noindent{ Table 2. Clipped Patch Data}\\
\begin{tabular}
{l l l l l l}
Name&$f(\%)$&$<N>$&$m_J(\sigma)$&$m_H(\sigma)$ &$m_{K_s}(\sigma)$ \cr
\hline
1&94 &475&$+21.2$ (56)&$+20.5$ (52)&$+20.1$ (36)\\
2&92&519&$+20.4$ (98)&$+20.1$ (75)&$+19.7$ (49)\\
3&93&482&$+21.3$ (48)&$+20.5$ (47)&$+20.1$ (32)\\
4&94&520&$+20.1$ (62)&$+20.3$ (65)&$+19.8$ (44)\\
5&93&489&$+21.3$ (46)&$+20.5$ (55)&$+20.1$ (33)\\
6&92&509&$+20.9$ (70)&$+20.3$ (62)&$+20.1$ (39)\\
7&94&472&$+20.9$ (67)&$+20.2$ (65)&$+19.8$ (45)\\
\hline
\end{tabular}
\vskip 0.5truein
\newpage

\vskip 0.2truein
\noindent{ Table 3. Limits to Foreground Fluctuation Contributions}\\
\begin{tabular}
{l l l l l}
Component&[J-H]&[H-K]&$I_\nu (K_s)$ (MJy/sr)&K-band Fluctuation (\nwm2sr )\cr
\hline
Atmosphere + Instrument&+1.9&+1.2&295& $< $2 \\
Zodiacal Light&+0.4&-0.3&0.1&$<$ 1  \\
Cirrus/DISM&-0.6&-1.1&0.2& 1 \\
Diffuse Star Light&+0.5&+0.7&0.2& $\lt$ 9 \\
\hline
\end{tabular}
\vskip 0.5truein
\newpage

Figure 1: Histogram of the background noise dispersion for all of the
2MASS images employed in this analysis for J-band (solid), H-band (dotted)
and K$_s$ band (dashed).

Figure 2: Histogram of the number of co-adds in each pixel
for the final image in K$_s$ band.

Figure 3: (Top) Individual 2MASS images for 7.8-second integration in J, H and
K$_s$ band. (Bottom) Corresponding deep-integration co-added images for an
effective integration time of 3,900 seconds.

Figure 4: Co-added calibration strip divided into seven patches as described in
the text. Actual sky image is shown in the left-hand columns. Distribution of
pixel samples employed to construct the averages in the integrated images shown
in the right-hand columns.

Figure 5: (a) Left: Color-color plot of the various emission components in Patch 3. The plotting symbols
indicate the following features: 
(X symbols)
indicate the color corresponding to the arc-second scale power 
spectrum amplitude
for all seven image patches.
(Ring) circumscribes the data for $10^5$ unresolved objects in the
2MASS Point Source Catalog.
(Diamond) The mean color of the stars photometered in Patch 3.
(Triangle) is the predicted zodiacal light emission. 
(Square)
indicates the mean color of the atmosphere obtained from an average of the 2080 images with  1-$\sigma$ error bars. 
(Asterisk) is the residual emission of the background of the co-added, 
clipped Patch 3 field. (b) Right: Color-magnitude plot of faint point sources in Patch 3.

Figure 6: Power spectrum of a single, 7.8-second image showing (lower curve)
the
influence of the beam and (top curve) the beam-subtracted spectrum.

Figure 7: Power spectrum of the K$_s$ band Patch 3 region showing: (top
curve) the power-law residual and (bottom points) the white-noise
contribution from atmospheric+instrument noise.

Figure 8: Power spectrum of deep-integration field in J, H and K$_s$ bands
showing the vertical and horizontal `artifact' caused by striping effects in
the original image.

Figure 9: Comparison of the seven patch sigmas in J (triangles), H
(squares)  and K$_s$ (crosses)
 bands before (horizontal axis) and after (vertical axis)
Fourier filtering to remove the striping
artifacts.

Figure 10: Power spectra of each of seven patches in each band. The value
for the background sigma employed in clipping point sources is indicated
in each panel.

Figure 11: Variation of the background noise dispersion for each image
as a function of airmass. The dispersion is given in units of \nwm2sr .

Figure 12: Distribution of the background pixel intensities following clipping
for J, H and K$_s$ bands in Patch 3.

Figure 13: Image of clipped field for Patch 3 in K$_s$ band to show the
locations of clipped pixels and the smoothness of the remaining background
emission. The gray scaling range extends from -5 to +5 \nwm2sr .

Figure 14: Power spectrum of the A-B Patch 3 field in J, H and K$_s$ bands.

Figure 15: 
Shows the power spectrum for the unclipped Patch 5 (diamonds) and its 
residual power spectrum after clipping and destriping (plus signs). This is 
juxtaposed against a test field with 7.8 sec exposure in the Galactic plane 
located at Galactic $(l, b)= (358.9^\circ, -21.6^\circ)$ which contains a comparable number of stars to that in Patch 5 
at the final co-add. The fraction of the pixels clipped in the former is 
about 14\%. The power spectrum for unlicpped test field is shown with open 
triangles; it is scaled to the unclipped power spectrum for Patch 5. The 
power spectrum for the clipped test field is shown with dots with the error 
bars.

Figure 16: 
Cumulative point source counts, N(m), in Patch 3 (triangles). The vertical 
axis is in units of Log(stars/deg$^2$). Error bars reflect the 
statistical uncertainty in the number of counted stars in Patch 3 in each 
magnitude bin. Also shown (solid line) is the predicted number of stars based
on the DIRBE Faint Source Model, and the expected number of galaxies 
(dashed line) according to a summary by Djorgovski et al. (1995).

Figure 17: (a) Left: Histogram of the rms radii of the discrete sources identified by SExtract in Patch 3
in K$_s$ band. (b) Right: Variation of the angular sizes of the discrete sources identified by
SExtract, versus their integrated K$_s$ magnitude.

\newpage
\clearpage
\begin{figure}
\centering
\leavevmode
\epsfxsize=1.0
\columnwidth
\epsfbox{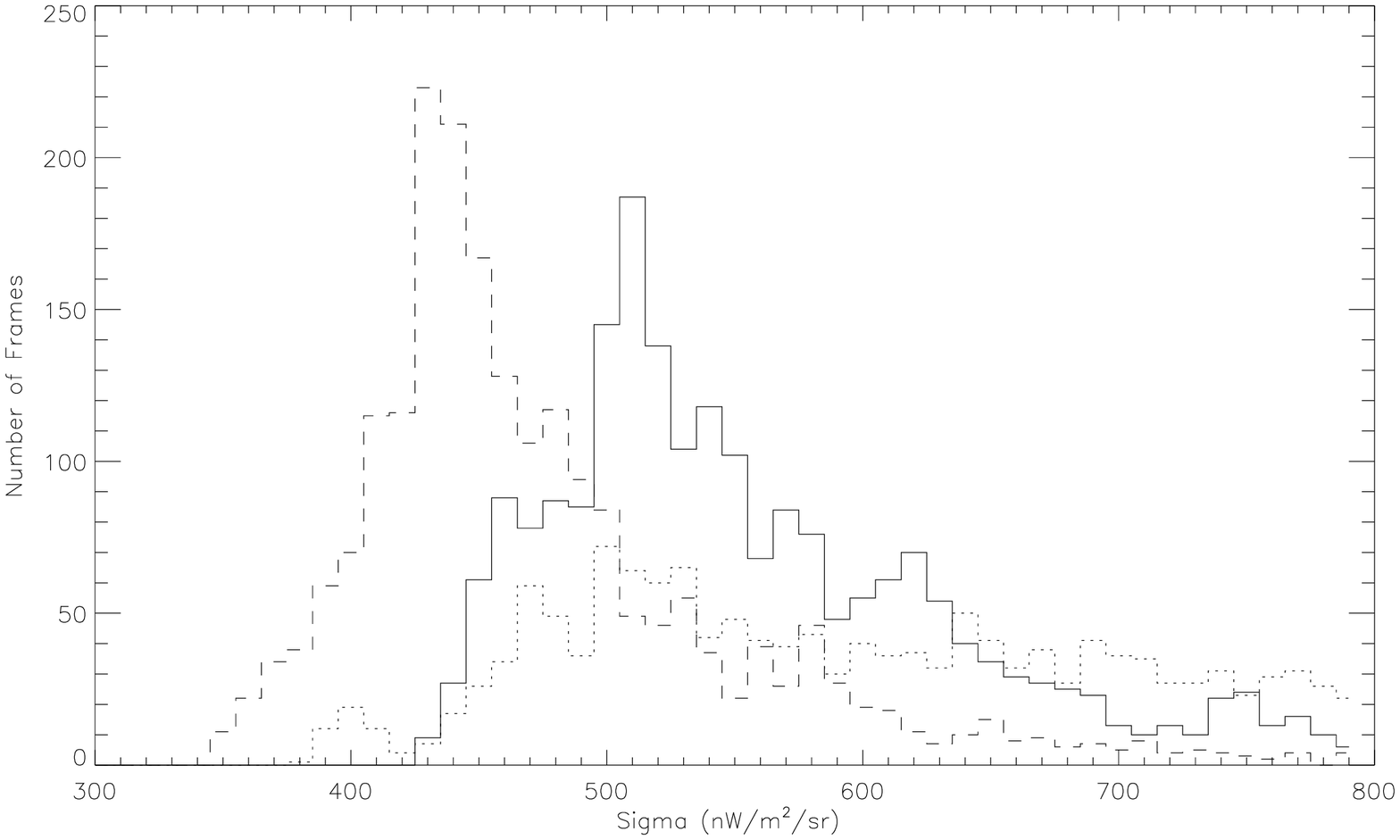}
\caption[]{Figure 1} 
\end{figure}

\clearpage
\begin{figure}
\centering
\leavevmode
\epsfxsize=1.0
\columnwidth
\epsfbox{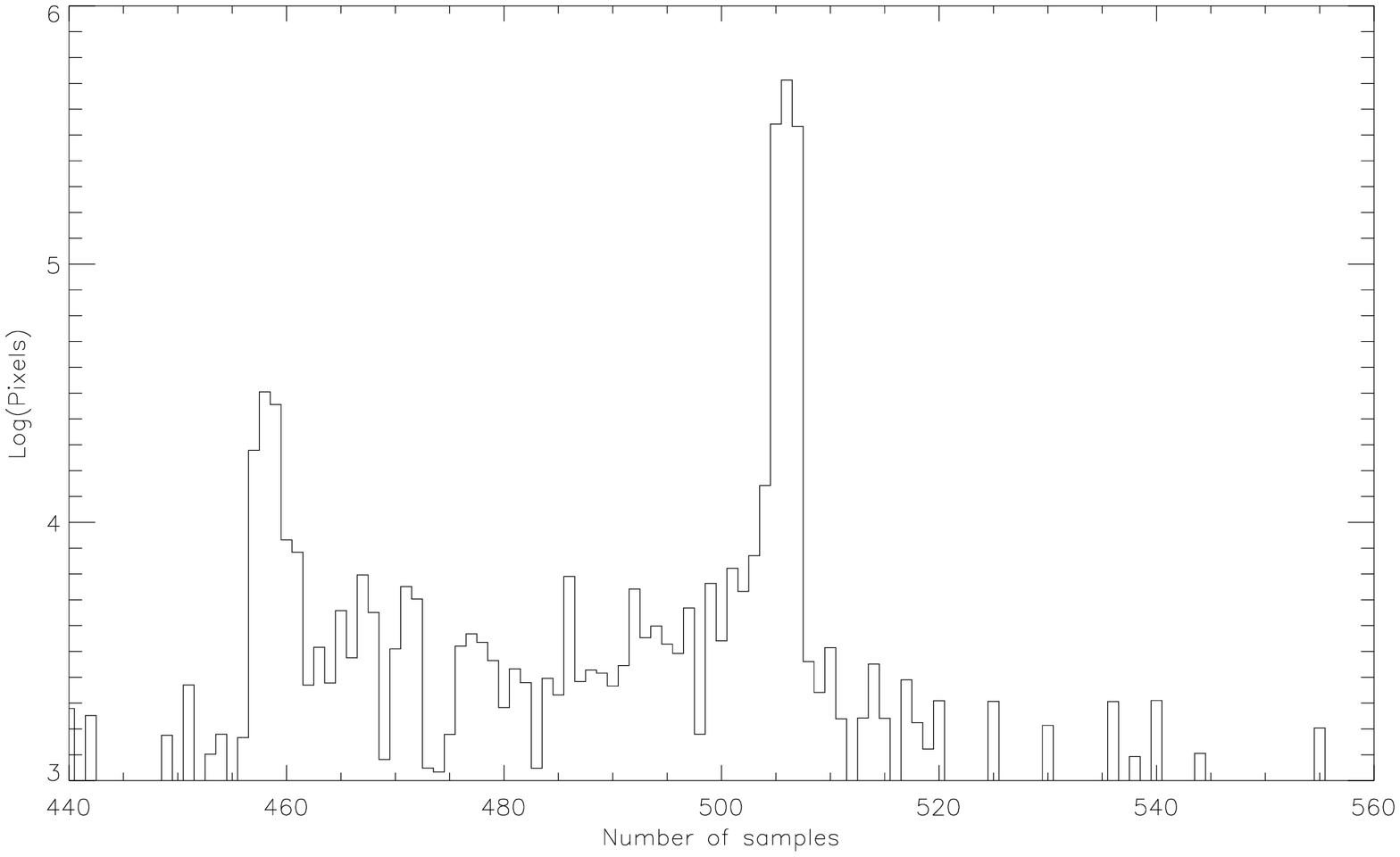}
\caption[]{Figure 2}
\end{figure}

\clearpage
\begin{figure}
\centering
\leavevmode
\epsfxsize=1.0
\columnwidth
\epsfbox{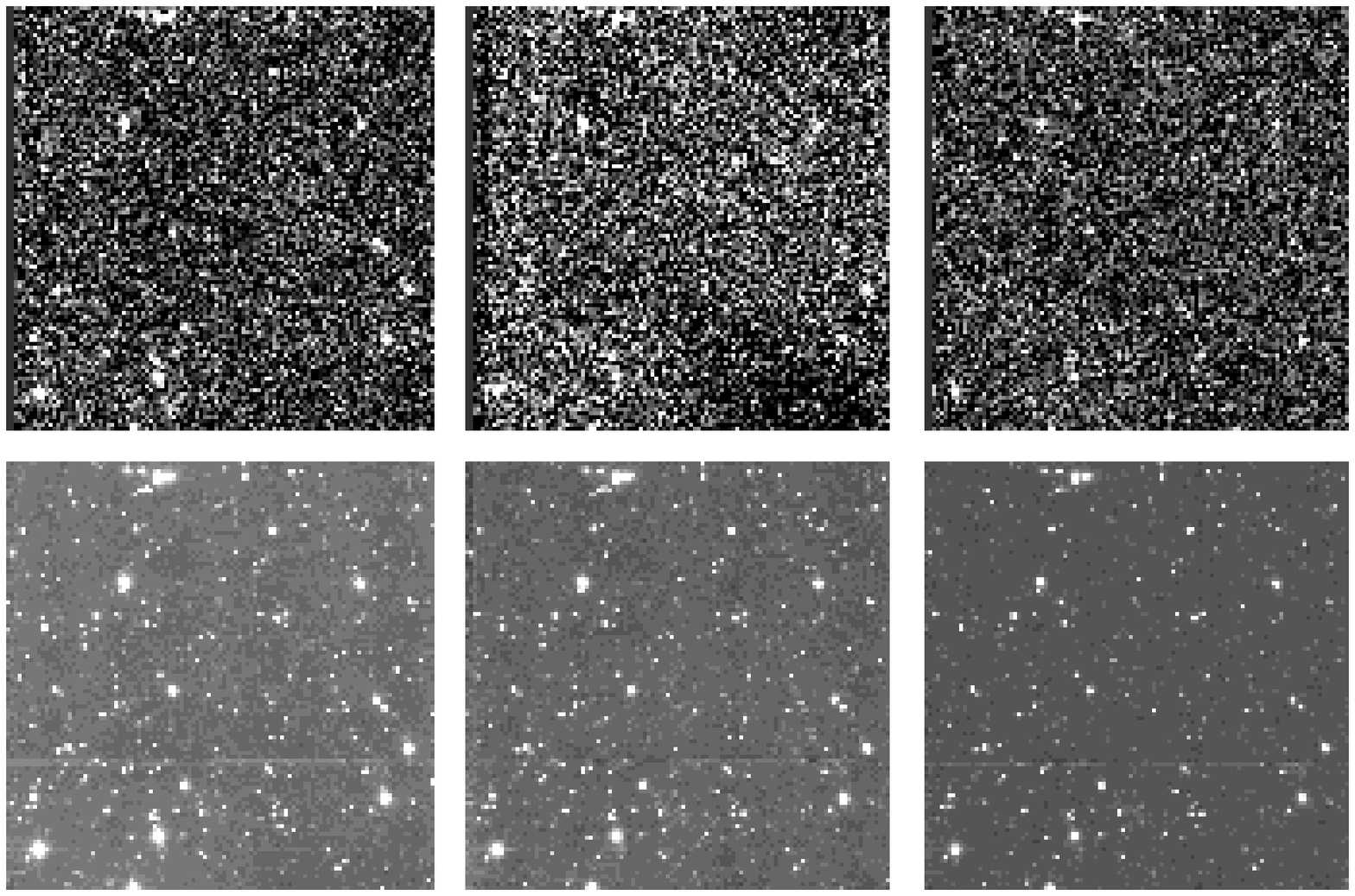}
\caption[]{Figure 3} 
\end{figure}

\clearpage
\begin{figure}
\centering
\leavevmode
\epsfxsize=1.0
\columnwidth
\epsfbox{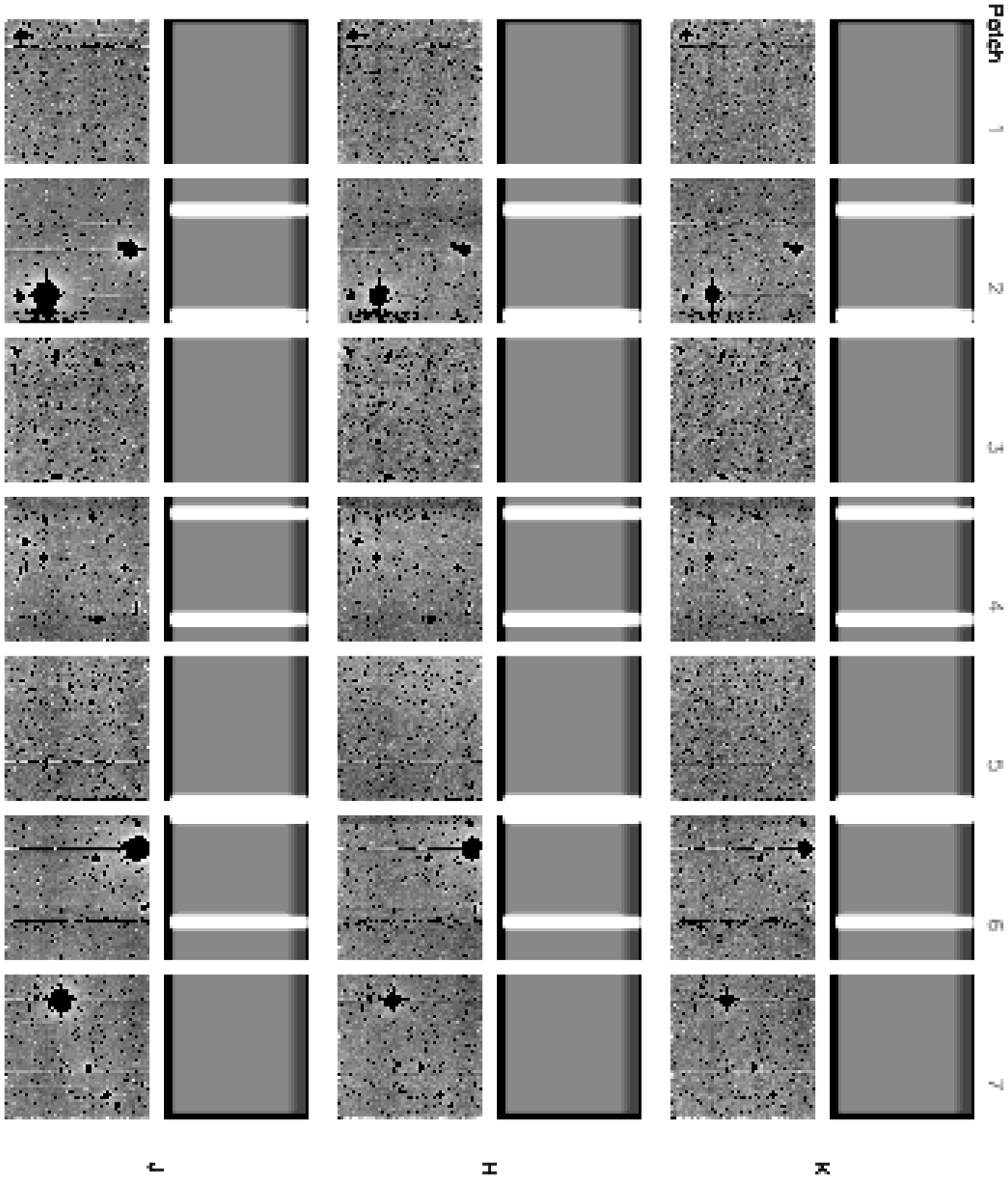}
\caption[]{Figure 4}
\end{figure}

\clearpage
\begin{figure}
\centering
\leavevmode
\epsfxsize=1.0
\columnwidth
\epsfbox{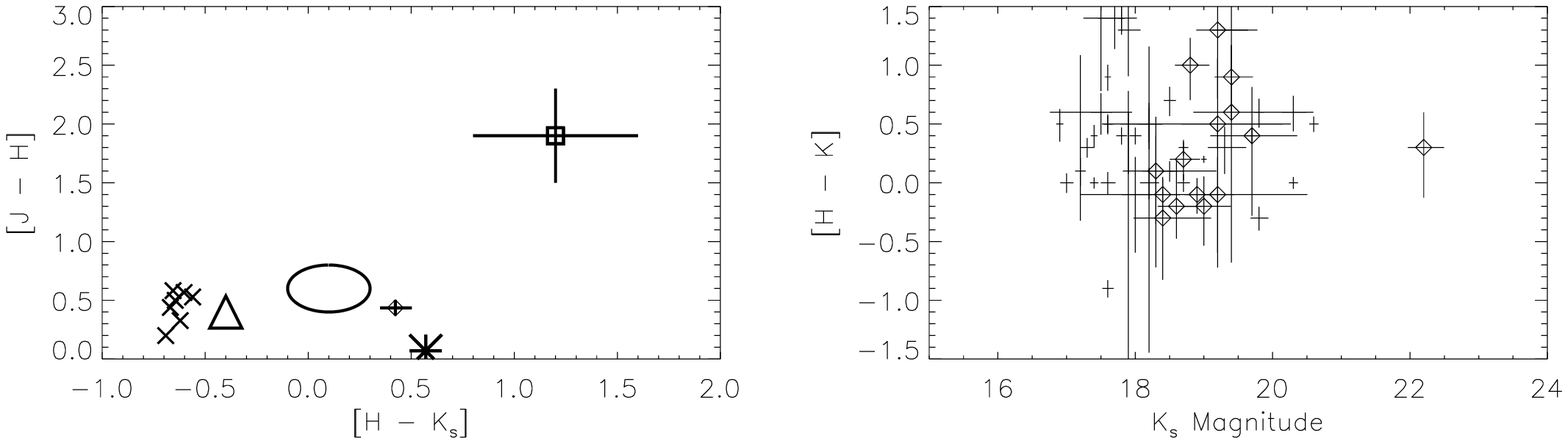}
\caption[]{Figure 5}
\end{figure}

\clearpage
\begin{figure}
\centering
\leavevmode
\epsfxsize=1.0
\columnwidth
\epsfbox{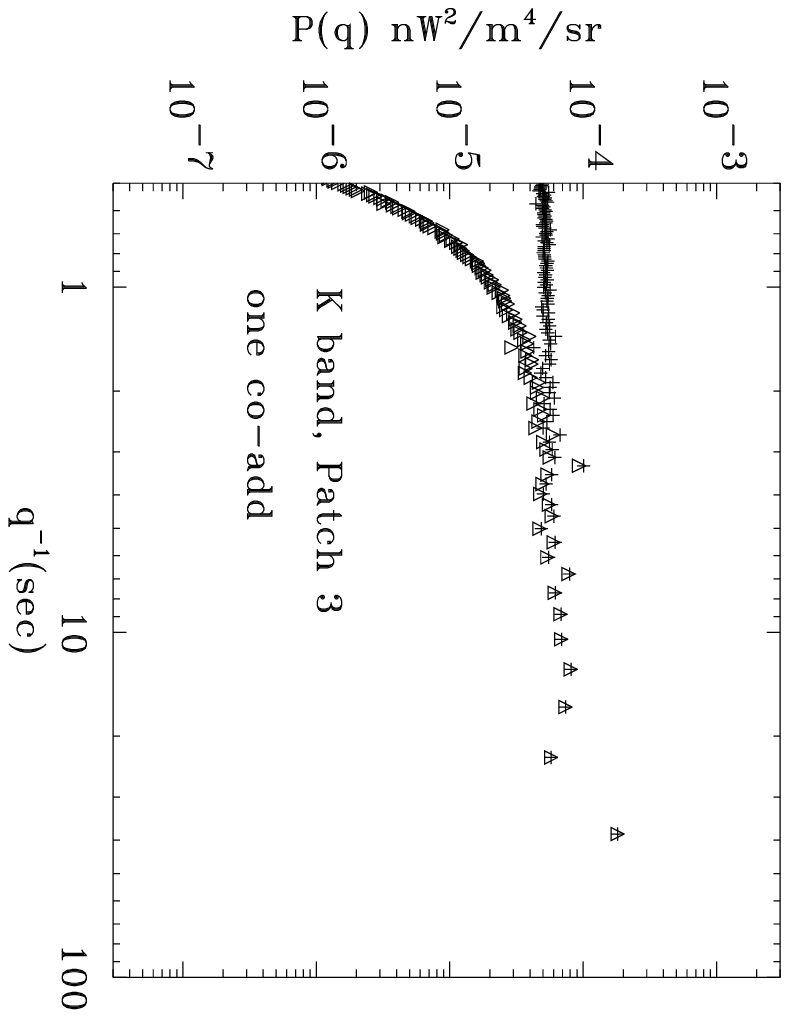}
\caption[]{Figure 6}
\end{figure}

\clearpage
\begin{figure}
\centering
\leavevmode
\epsfxsize=1.0
\columnwidth
\epsfbox{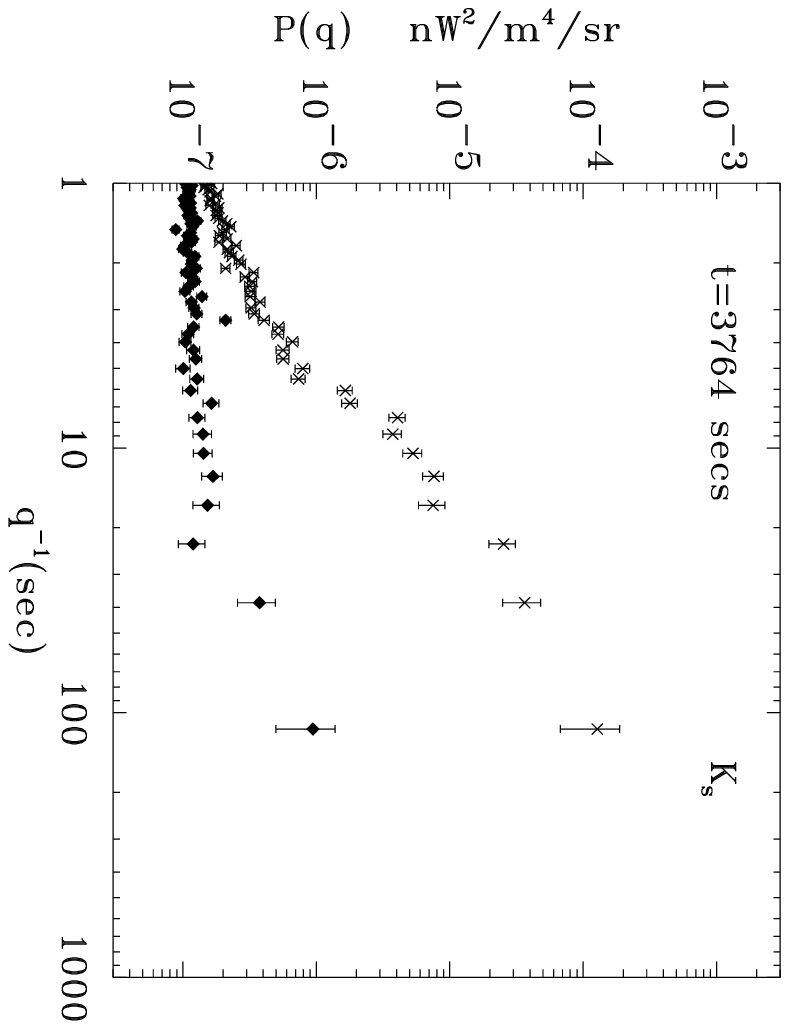}
\caption[]{Figure 7}
\end{figure}

\clearpage
\begin{figure}
\centering
\leavevmode
\epsfxsize=1.0
\columnwidth
\epsfbox{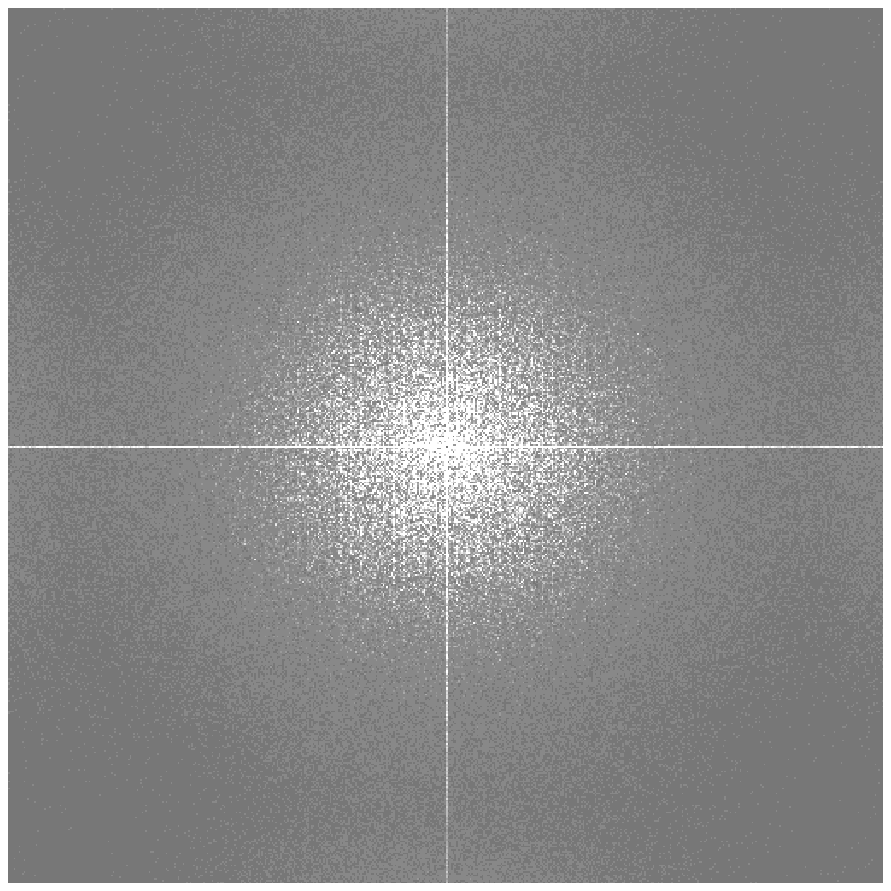}
\caption[]{Figure 8}
\end{figure}

\clearpage
\begin{figure}
\centering
\leavevmode
\epsfxsize=1.0
\columnwidth
\epsfbox{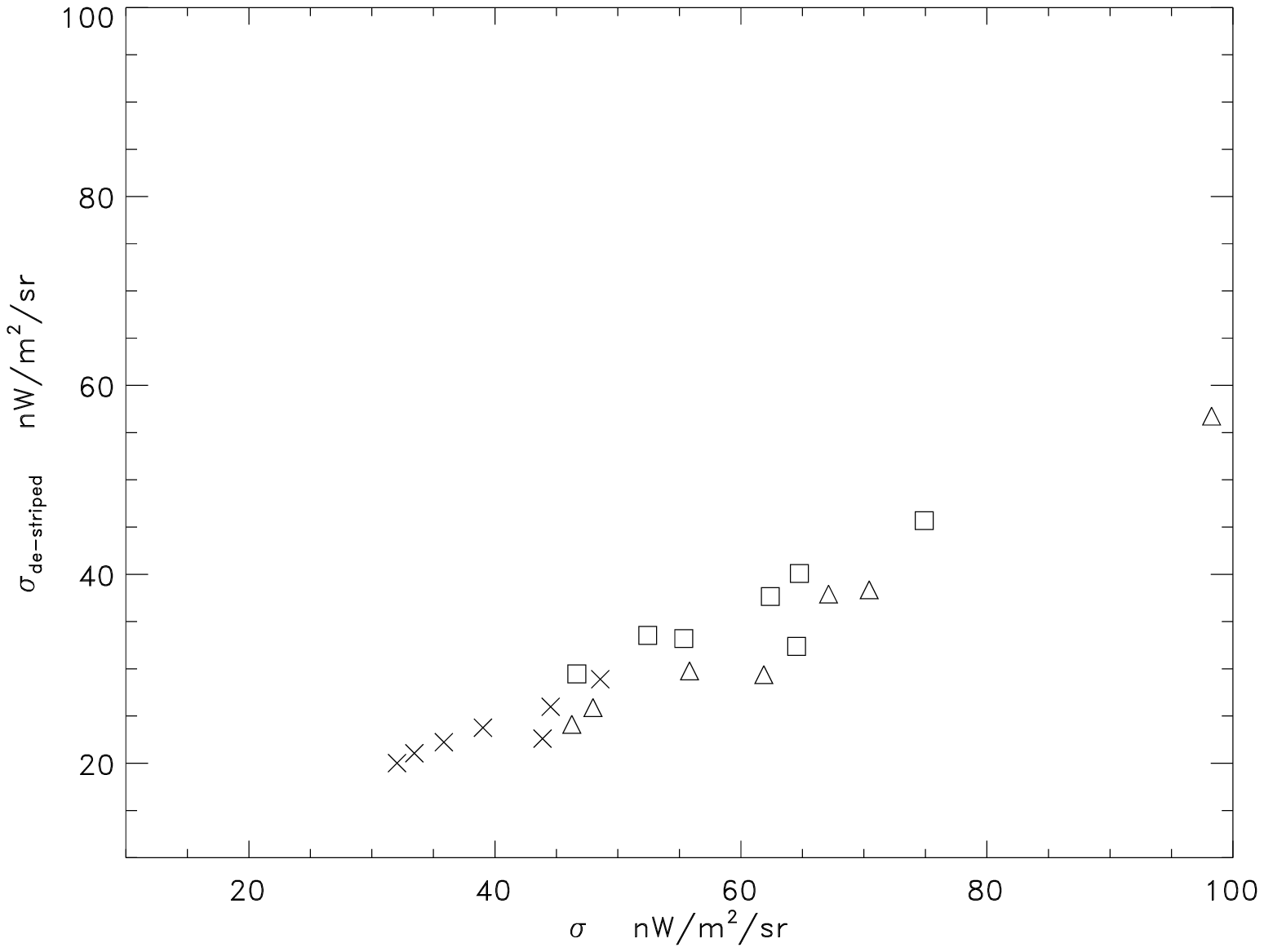}
\caption[]{Figure 9}
\end{figure}

\clearpage
\begin{figure}
\centering
\leavevmode
\epsfxsize=1.0
\columnwidth
\epsfbox{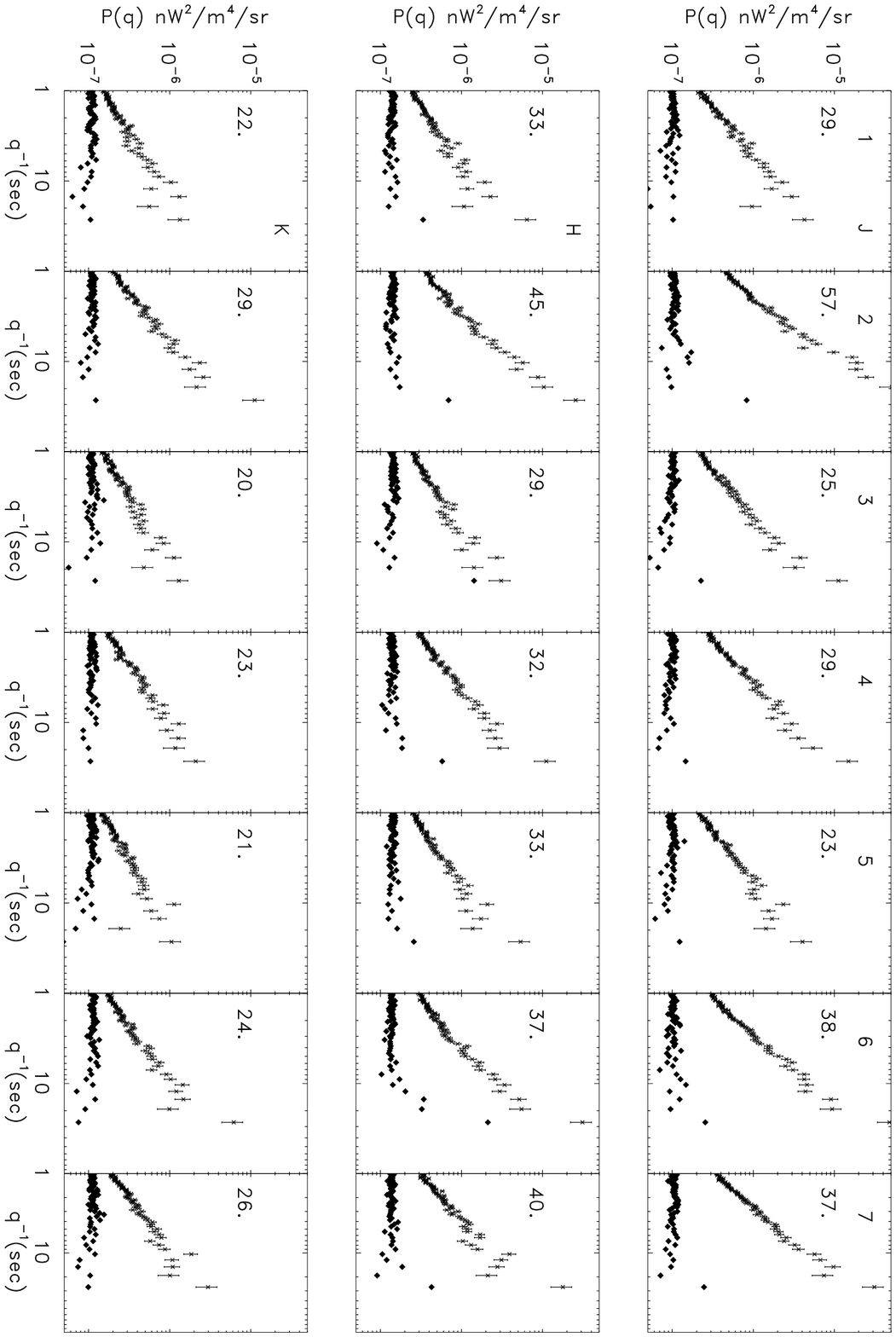}
\caption[]{Figure 10}
\end{figure}

\clearpage
\begin{figure}
\centering
\leavevmode
\epsfxsize=1.0
\columnwidth
\epsfbox{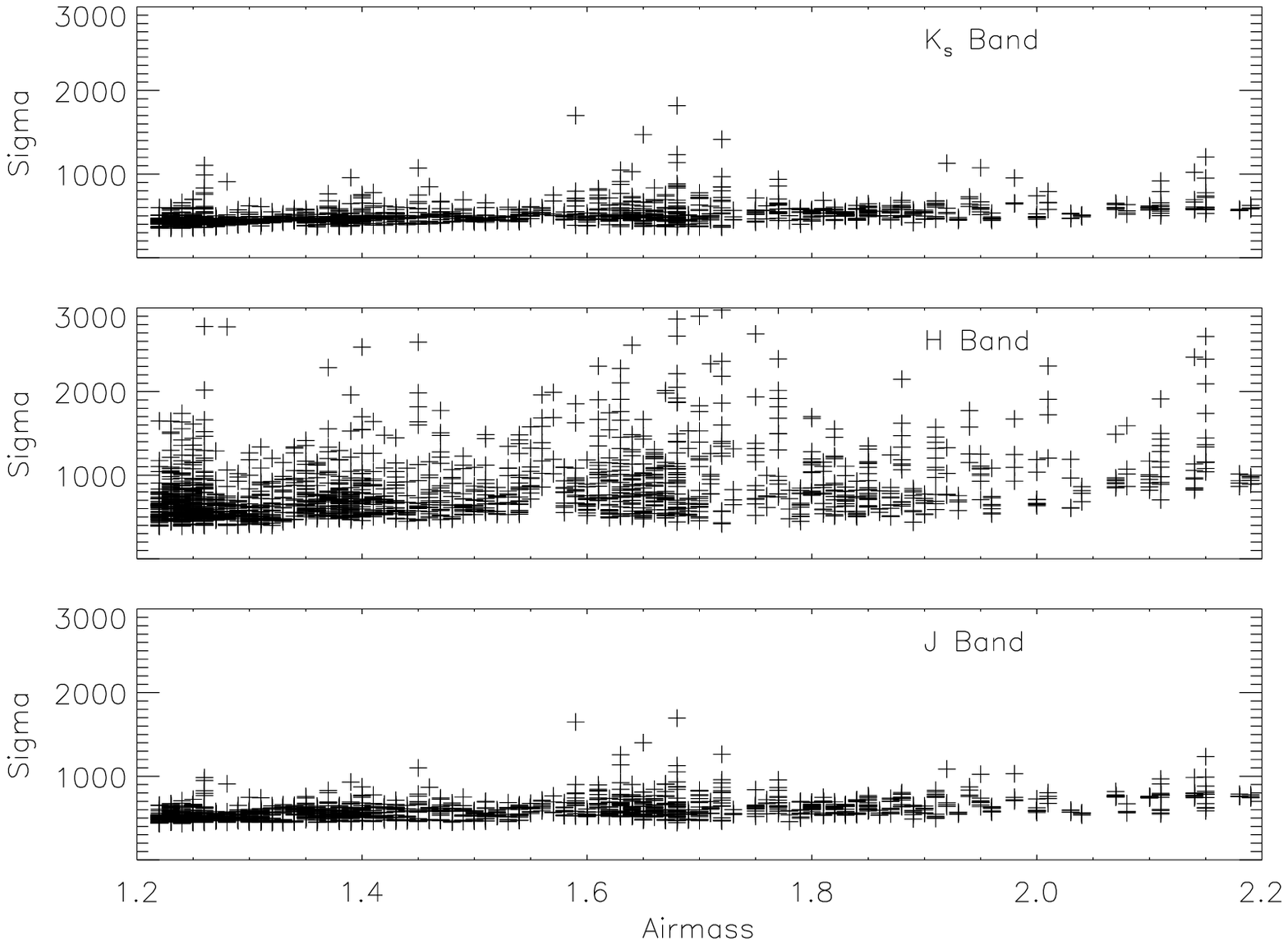}
\caption[]{Figure 11}
\end{figure}

\clearpage
\begin{figure}
\centering
\leavevmode
\epsfxsize=1.0
\columnwidth
\epsfbox{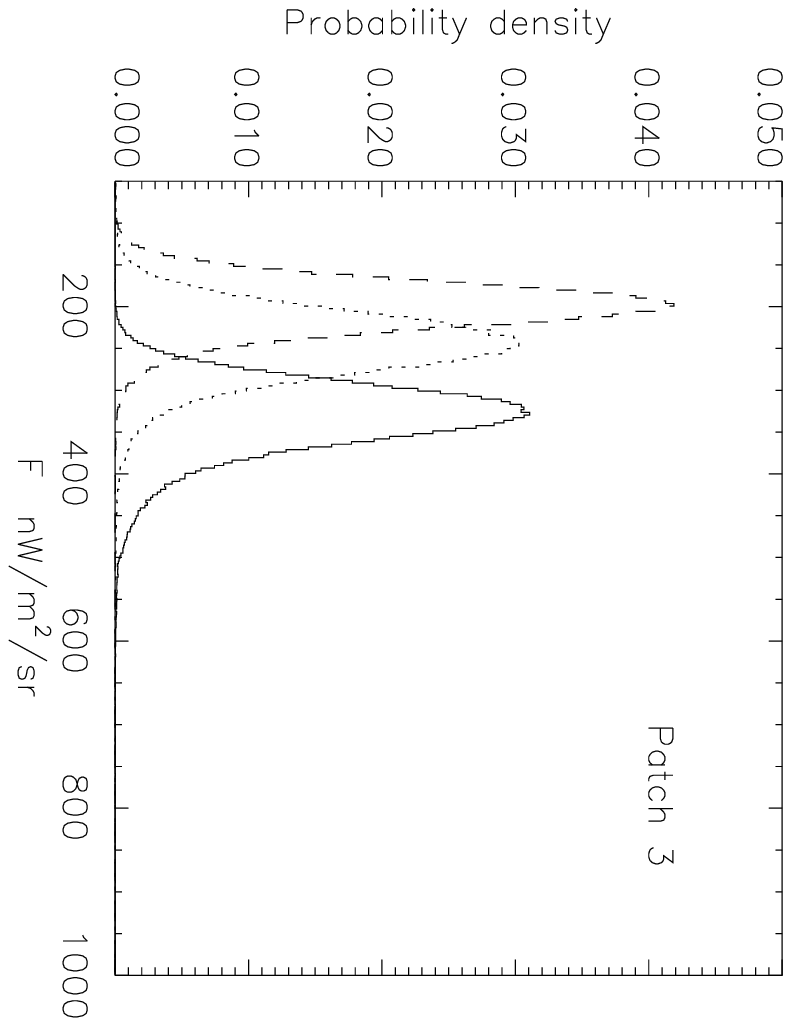}
\caption[]{Figure 12}
\end{figure}

\clearpage
\begin{figure}
\centering
\leavevmode
\epsfxsize=1.0
\columnwidth
\epsfbox{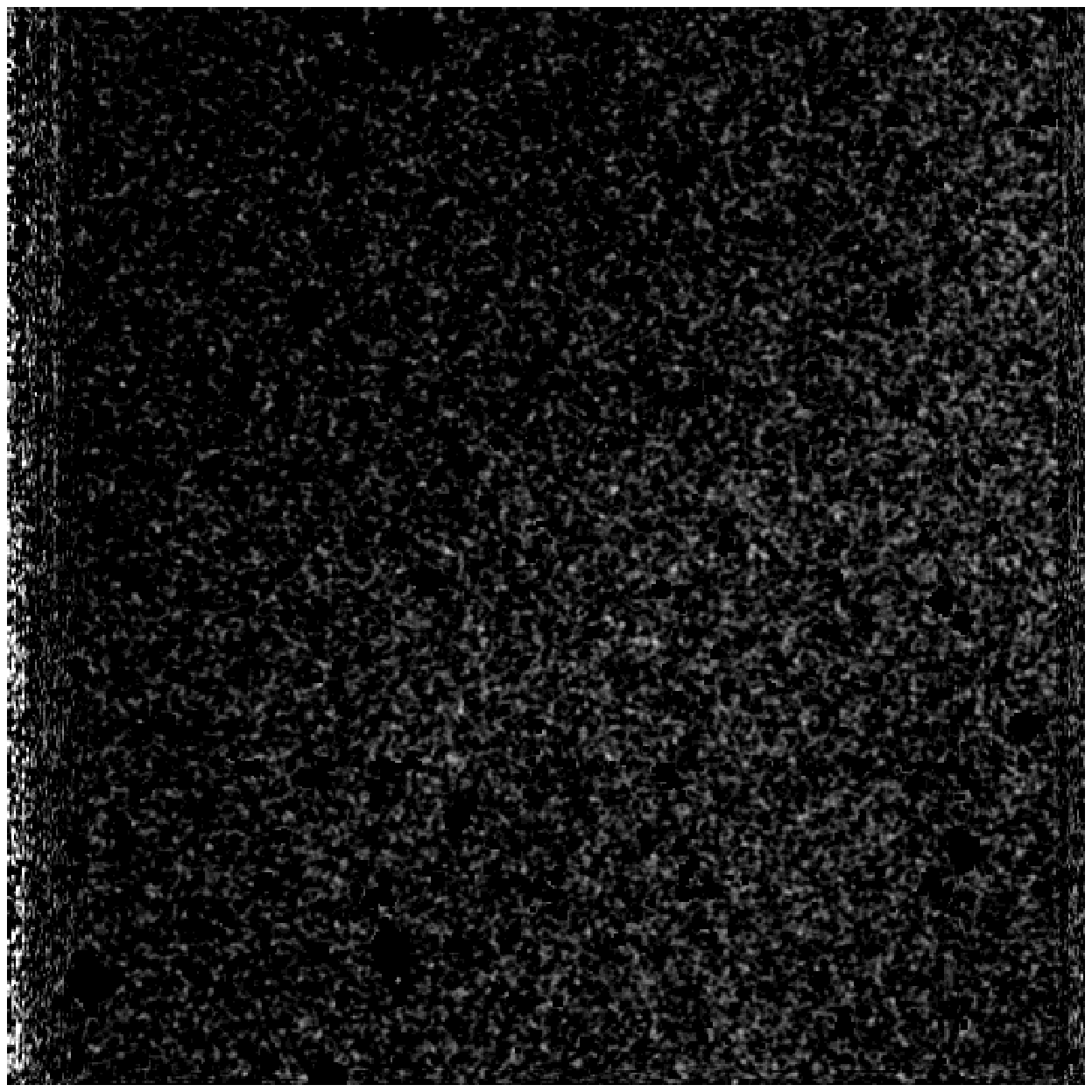}
\caption[]{Figure 13}
\end{figure}

\clearpage
\begin{figure}
\centering
\leavevmode
\epsfxsize=1.0
\columnwidth
\epsfbox{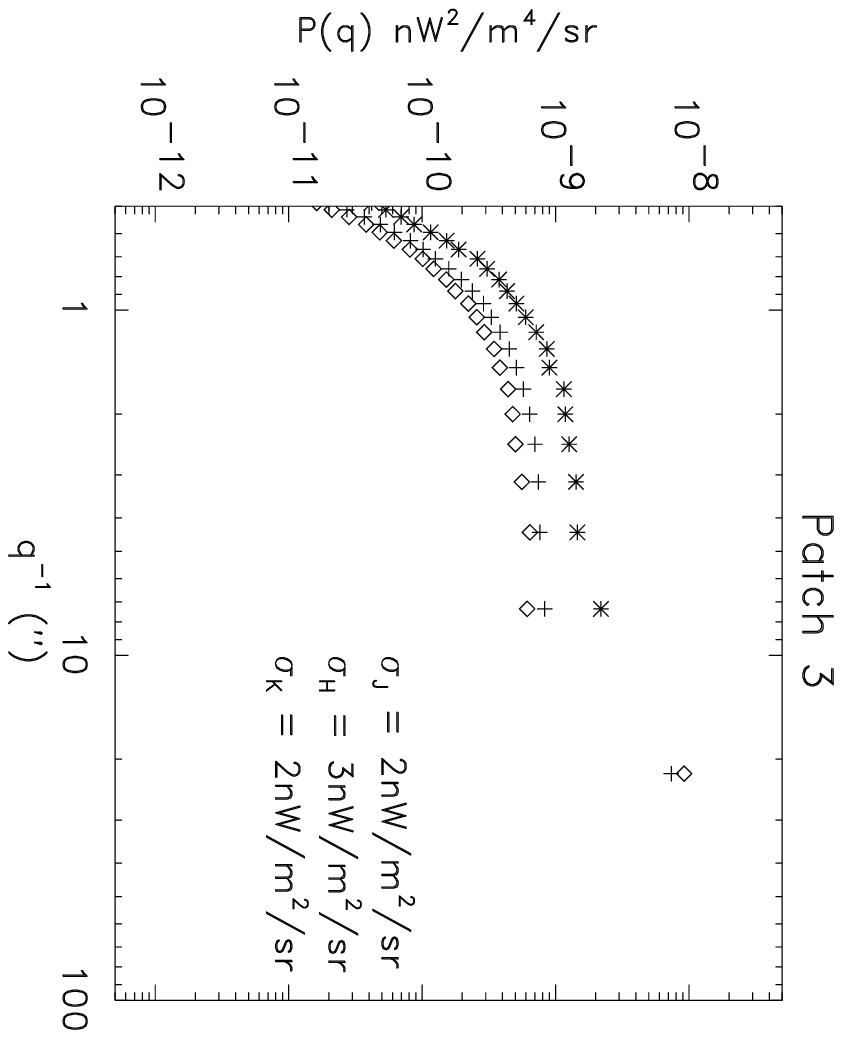}
\caption[]{Figure 14}
\end{figure}

\clearpage
\begin{figure}
\centering
\leavevmode
\epsfxsize=1.0
\columnwidth
\epsfbox{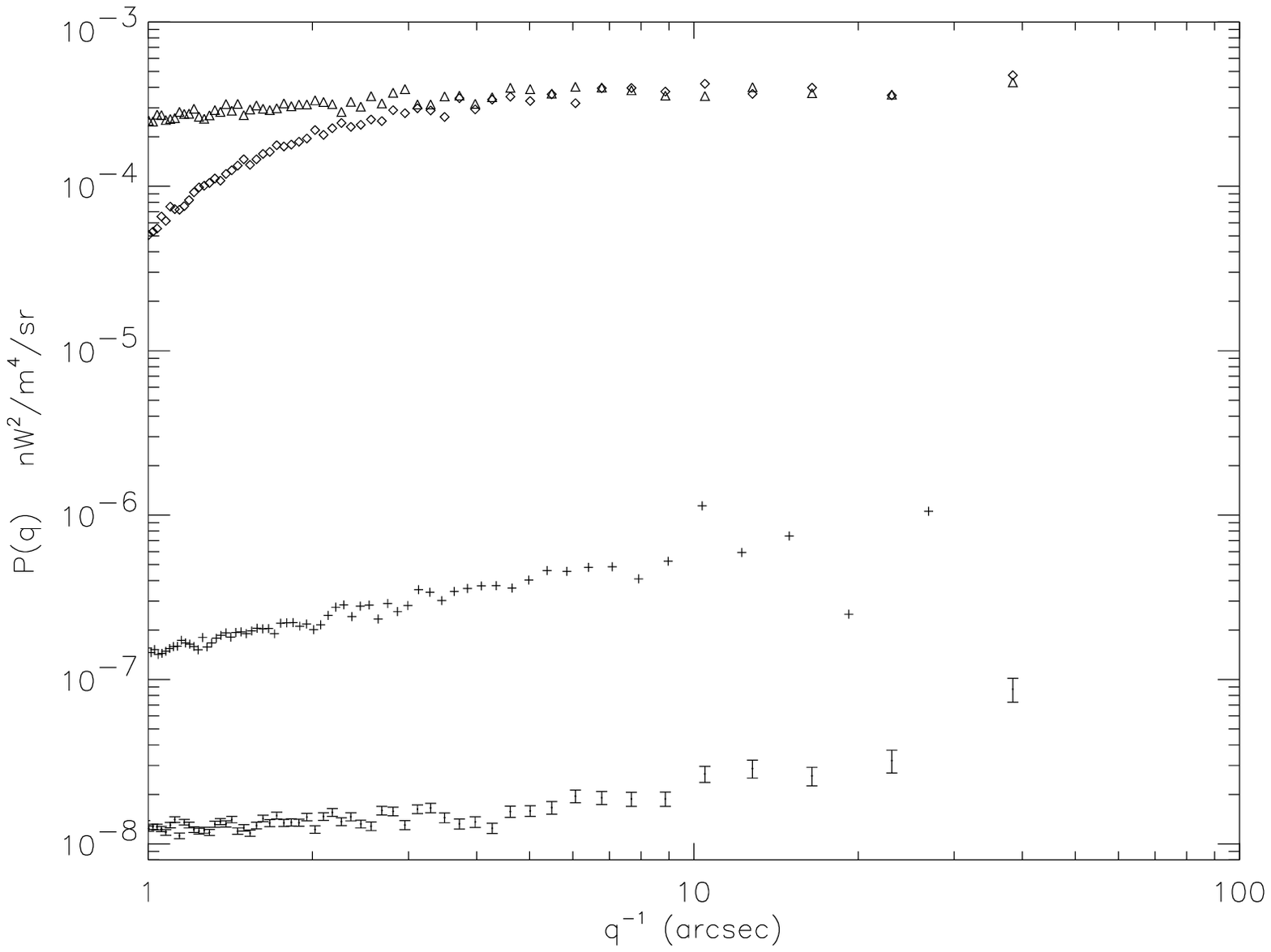}
\caption[]{Figure 15 }
\end{figure}

\clearpage
\begin{figure}
\centering
\leavevmode
\epsfxsize=1.0
\columnwidth
\epsfbox{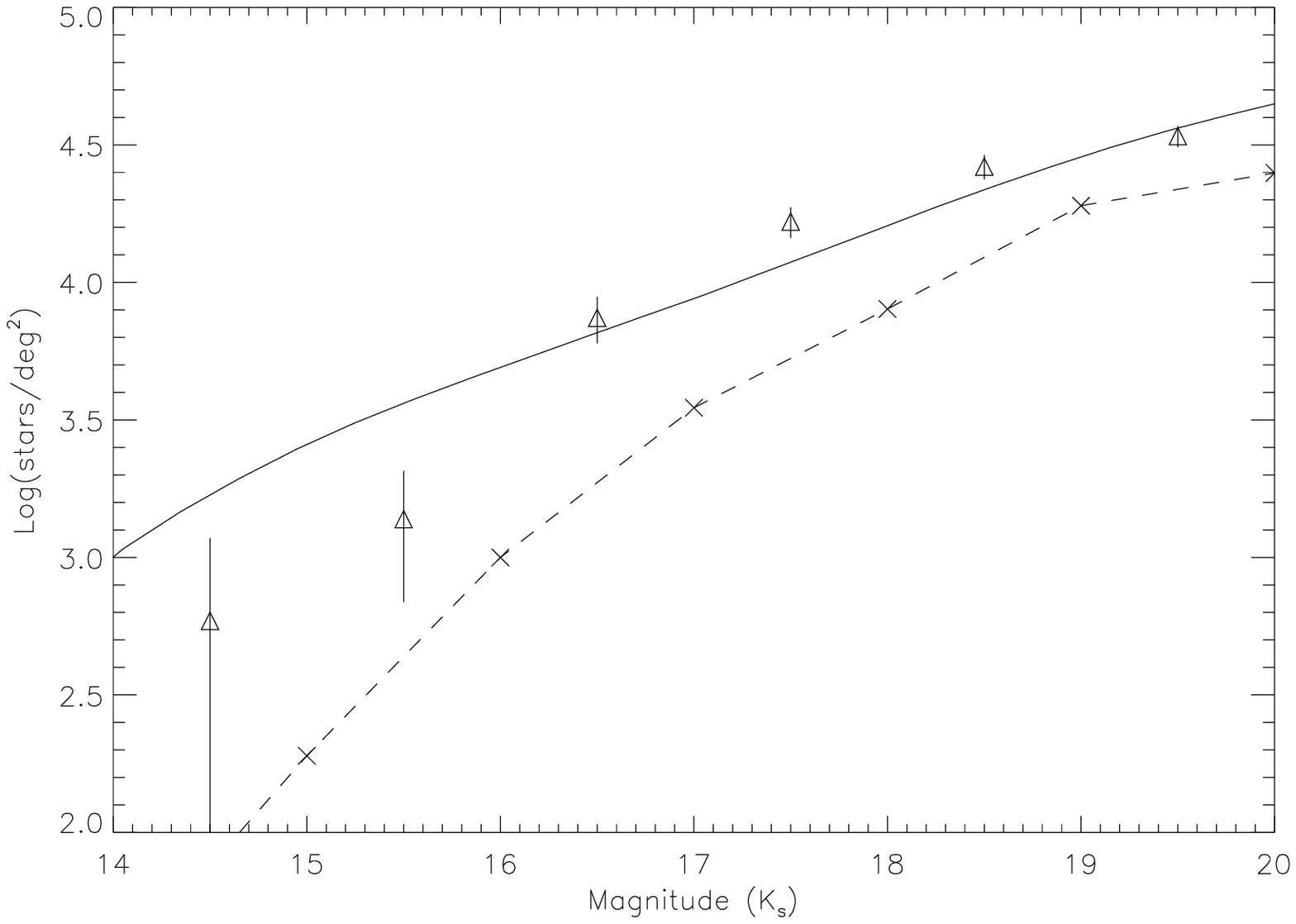}
\caption[]{Figure 16}
\end{figure}

\clearpage
\begin{figure}
\centering
\leavevmode
\epsfxsize=1.0
\columnwidth
\epsfbox{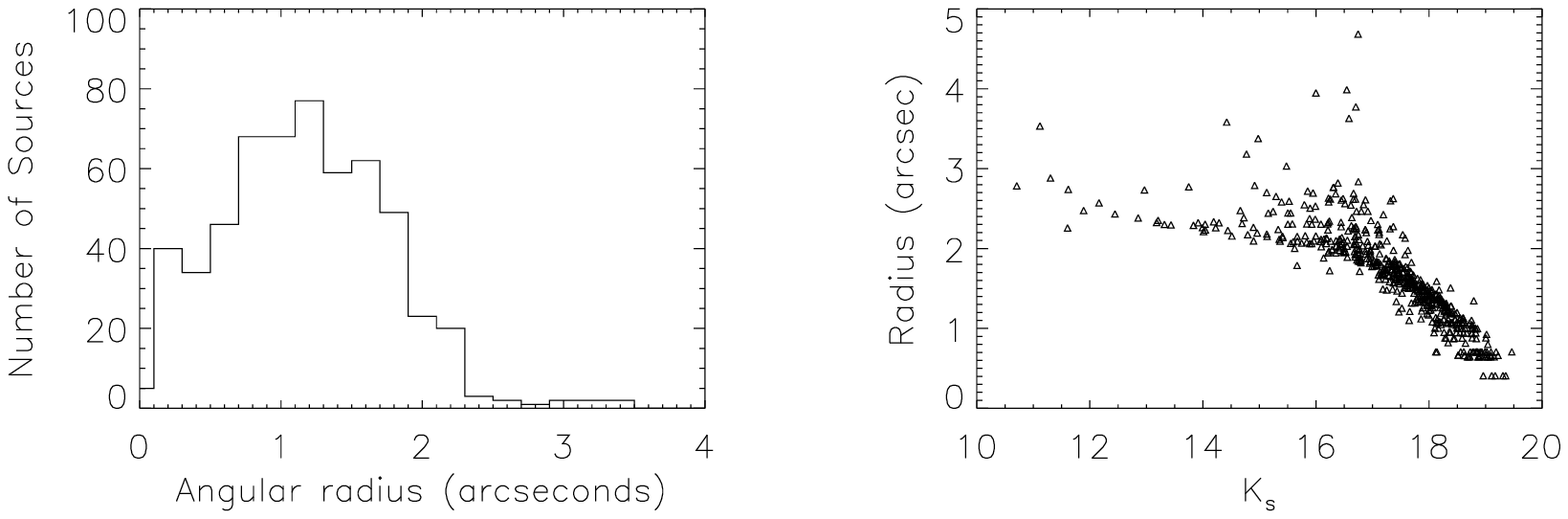}
\caption[]{Figure 17}
\end{figure}

\end{document}